\documentclass[letter]{aa}
\nolinenumbers
\usepackage{graphicx}
\usepackage{txfonts}
\usepackage{hyperref}
\usepackage{natbib}
\usepackage{booktabs}
\usepackage{siunitx}
\usepackage{placeins}

\bibpunct{(}{)}{;}{a}{}{,}
\nolinenumbers
\newcommand{\Konus}{\textit{Konus}/Wind }

\newcommand{\GBM}{\textit{Fermi}/GBM}

\defcitealias{Maccary26}{M26}

\begin{document}
\nolinenumbers
    \title{The hidden population of long gamma-ray bursts from compact object mergers} 

    \titlerunning{A new set of long gamma-ray burst merger candidates}


\author{R.~Maccary\thanks{mccrnl@unife.it}\inst{\ref{unife},\ref{inafoas}}
   \and C.~Guidorzi\inst{\ref{unife},\ref{inafoas},\ref{infnfe}}
   \and L.~Amati\inst{\ref{inafoas}}
   \and M.~Bulla\inst{\ref{unife},\ref{infnfe},\ref{inaf_te}}
   \and S.~Kobayashi\inst{{\ref{ljmu}}}
   \and M.~Maistrello\inst{\ref{unife},\ref{inafoas}}
   \and A.~Rossi\inst{\ref{inafoas}}
   \and G.~Stratta\inst{\ref{inafoas},\ref{infn_bo}}
   \and A.~Tsvetkova\inst{\ref{inaf_ca},\ref{ioffe}}}
            
    \institute{Department of Physics and Earth Science, University of Ferrara, Via Saragat 1, I-44122 Ferrara, Italy
            \label{unife}
        \and 
            INAF –- Osservatorio di Astrofisica e Scienza dello Spazio di Bologna, Via Piero Gobetti 101, 40129 Bologna, Italy
            \label{inafoas}
        \and 
            INFN –- Sezione di Ferrara, Via Saragat 1, 44122 Ferrara, Italy 
            \label{infnfe}
        \and
            INAF –- Osservatorio Astronomico d’Abruzzo, Via Mentore Maggini snc, Teramo, 64100, Teramo, Italy
            \label{inaf_te}
        \and 
           Astrophysics Research Institute, LJMU, IC2, Liverpool Science Park, 146 Brownlow Hill, Liverpool L3 5RF, UK
           \label{ljmu}
        \and 
            INFN –- Sezione di Bologna Viale C. Berti Pichat 6/2 - 40126 Bologna, Italy
            \label{infn_bo}
        \and
            Dipartimento di Fisica, Università degli Studi di Cagliari, SP Monserrato-Sestu, km 0.7, I-09042 Monserrato, Italy
            \label{inaf_ca}
        \and
            Ioffe Institute, Politekhnicheskaya 26, 194021 St. Petersburg, Russia
            \label{ioffe}
            }
    
    \date{Received date / Accepted date }
    
    \abstract
    {The prompt-emission time profiles of GRB\,230307A and other long-duration compact object merger (COM) candidates exhibit a unique set of temporal properties, characterised by a deterministic evolution of waiting times and pulse widths.}
    {We searched the \textit{Fermi}/GBM catalogue for other unidentified long COM candidates exhibiting temporal properties similar to those observed in GRB\,230307A.}
    {We examined the temporal and spectral prompt-emission properties of GRBs featuring at least eight light-curve peaks. For candidates, all with unknown redshifts, that exhibited properties similar to GRB\,230307A, we analysed their trajectories in the $E_{\rm p,i}$--$E_{\rm iso}$ plane as a function of redshift. We then evaluated the joint likelihood of their compatibility with the $E_{\rm p,i}$--$E_{\rm iso}$ relation satisfied by the bulk of long GRBs. Furthermore, we calculated their minimum variability timescales (MVTs) for comparison against known COM and collapsar populations.}
    {We identified 9 COM candidates with unknown redshifts and demonstrated that there are at least two outliers of the $E_{\rm p,i}$--$E_{\rm iso}$ relation with $3.1 \sigma$ (Gaussian) confidence level. Furthermore, their MVTs are more consistent with those of COM than with collapsar GRBs.}
    {These results indicate that this specific set of temporal properties can serve as a diagnostic tool to distinguish long-duration COMs from the broader collapsar population. Furthermore, our findings suggest that the fraction of unidentified COMs among long GRBs may be larger than previously assumed.}

    \keywords{(Stars:) $\gamma$-ray burst: general -- Methods: statistical}
    \maketitle
    \nolinenumbers
\section{Introduction}
\label{sec:intro}
Compact object mergers (COMs), typically associated with short $\gamma$-ray bursts (SGRBs), are sometimes also found in association with long duration GRBs (LGRBs), as was the case for GRB 230307A, a long burst accompanied by kilonova emission characteristic of COMs (e.g., \citealt{Dai24, Zhong24, Levan24, Yang24, Dalessi25, Gillanders25}). For this reason, GRBs are now often classified according to their progenitor and referred to as type-I and type-II GRBs, corresponding to COMs and massive star collapse, respectively, rather than being classified solely by their duration (e.g., \citealt{Zhang06_nat}).

\citet[hereafter M26]{Maccary26} identified a set of four distinctive properties characterising the evolution of the time profile of GRB\,230307A and of other COM candidates like GRB\,211211A and GRB\,060614. By contrast, the same properties are not seen in any typical long GRB associated with a type Ic-BL supernova (SN). The $E_{\rm p,i}-E_{\rm iso}$ relation (e.g., \citealt{Amati02,Amati06}) holds for type-II GRBs, while type-I lie above them in this plane and form a distinct cluster of points, thus offering a way to distinguish GRB progenitors. Moreover, a short ($\lesssim 100$~ms) minimum variability timescale (MVT) characterises the time profiles of type-I GRBs, both short and long \citep{MacLachlan12,MacLachlan13, Golkhou15,Camisasca23,Veres23,Maccary25,Stratta25}.
\newline \hspace*{0.5cm}The goal of this paper is to identify additional COM candidates within the \textit{Fermi}/GBM catalogue and to use the $E_{\rm p,i}-E_{\rm iso}$ relation to further investigate their origin along with the clues given by the MVT. In Sec. \ref{sec:data}, we select a sample of long type-I GRB candidates that show similar properties as those identified in \citetalias{Maccary26}. These GRBs being not associated with a redshift measurement, we study their tracks in the $E_{\rm p,i}-E_{\rm iso}$ plane as a function of the (unknown) redshift, and perform statistical tests to assess their compatibility with the $E_{\rm p,i}-E_{\rm iso}$ of type-II GRBs. In Sec. \ref{sec:discussion}, we discuss our results and present our conclusions. Hereafter, we use the flat-$\Lambda$CDM cosmology model with the latest cosmological parameter values $H_0 = 67.66$~km~Mpc$^{-1}$~s$^{-1}$ and $\Omega_0 = 0.31$ \citep{cosmoPlanck20}.

\section{Data analysis}
\label{sec:data}

\subsection{Sample selection}
\label{sec:sample_sel}
We inspected the bursts detected by \GBM, from launch to June 2024, looking for a behaviour similar to that observed in GRB\,230307A \citepalias{Maccary26}. The background subtraction was carried out as in \citet{Maccary25}, from which most GRBs were taken for the present analysis. The peaks present in the light curve (LC) were identified with {\sc mepsa} \citep{Guidorzi15a}. We selected the bursts showing at least eight statistically significant peaks in their LCs to ensure an adequate sampling of the temporal profile, enabling a statistically meaningful characterisation of the temporal evolution of the properties identified in \citetalias{Maccary26}. We ended up with 263 bursts.
We identified 9 type-I candidates (hereafter called Sample $S_0$) featuring pulses, whose peak rates (PRs), waiting times (WTs), and full widths at half maximum (FWHMs) clearly show a deterministic evolution over time, in line with what was observed in GRB\,230307A. They are reported in Table~\ref{tab:grb_type_I_candidates}. Three cases are showcased in Fig.~\ref{fig:evolution_three}, while the other six are reported in Fig.~\ref{fig:evolution_six_others} of the Appendix. Table~\ref{tab:fit_results_3ev} reports the parameters obtained for the exponential models, defined in Appendix~\ref{sec:exp_fit}, that describe the temporal evolution of the properties listed above. In Appendix \ref{sec:corr_tests}, we performed a Pearson correlation test (reported in Fig.~\ref{sec:corr_tests}) to further assess the observed correlations between the aforementioned four properties and the peak times of the pulses observed in the time profile of these bursts.
\begin{table*}[h!]
\centering
\caption{Type-I GRB candidates identified in this work (referred to as sample $S_0)$. 
}
\label{tab:grb_type_I_candidates}
\begin{tabular}{lccccccc}
\toprule
GRB & Fermi ID &$T_{90}$  &$\rm FWHM_{\rm min} $ &Fluence& $E_{\text{p}}$ & $\mathcal{L}_*$  \\
& &(s) & (ms) & ($10^{-6}$ erg/cm$^2)$ & (keV) &  \\
\midrule
GRB\,090831 & bn090831317 & $39.4 \pm 0.6$ & $69^{+24}_{-18}$ & $9.44 \pm 0.07$ & $256 \pm 137$ & $0.08$  \\
GRB\,140306A & bn140306146 & $51.7 \pm 0.8$ & $78^{+27}_{-20}$ & $57.80 \pm 0.09$ & $1937 \pm 209$ & $4.84$ \\
GRB\,150510A & bn150510139 & $51.9 \pm 0.5$ & $84^{+29}_{-22}$ & $98.62 \pm 0.09$ & $1270 \pm 50$ & $2.29$ \\
GRB\,170527A& bn170527480 & $49.2 \pm 1.6$  & $140^{+48}_{-36}$ &  $84.30 \pm 0.03$ & $1095 \pm 47$ & $1.94$ \\
GRB\,190415A & bn190415173 & $52.5 \pm 0.8$ & $15^{+5}_{-4}$ & $60.13 \pm 0.03$ & $1834 \pm 126$ & $4.60$ \\
GRB\,211019A & bn211019250 & $47.4 \pm 0.6$ & $86^{+30}_{-22}$ & $107.06 \pm 0.05$ & $1169 \pm 40$ & $1.90$\\
GRB\,220408B & bn220408311 & $32.5 \pm 1.1$ & $110^{+38}_{-28}$ & $34.60 \pm 0.07$ & $225 \pm 5$ & $-0.38$\\
GRB\,221121A & bn221121274 & $40.7 \pm 1.6$ & $110^{+27}_{-37}$ &  $14.93 \pm 0.03$ & $889 \pm 57$ & $3.41$ \\ 
GRB\,230304B & bn230304608 & $34.8 \pm 0.6$ & $93^{+32}_{-24}$ & $37.62 \pm 0.03$ & $301 \pm 11$ & $-0.25$ \\
\bottomrule
\\
\end{tabular}
\end{table*}
\begin{figure*}
\centering
\includegraphics[width=0.9\linewidth]{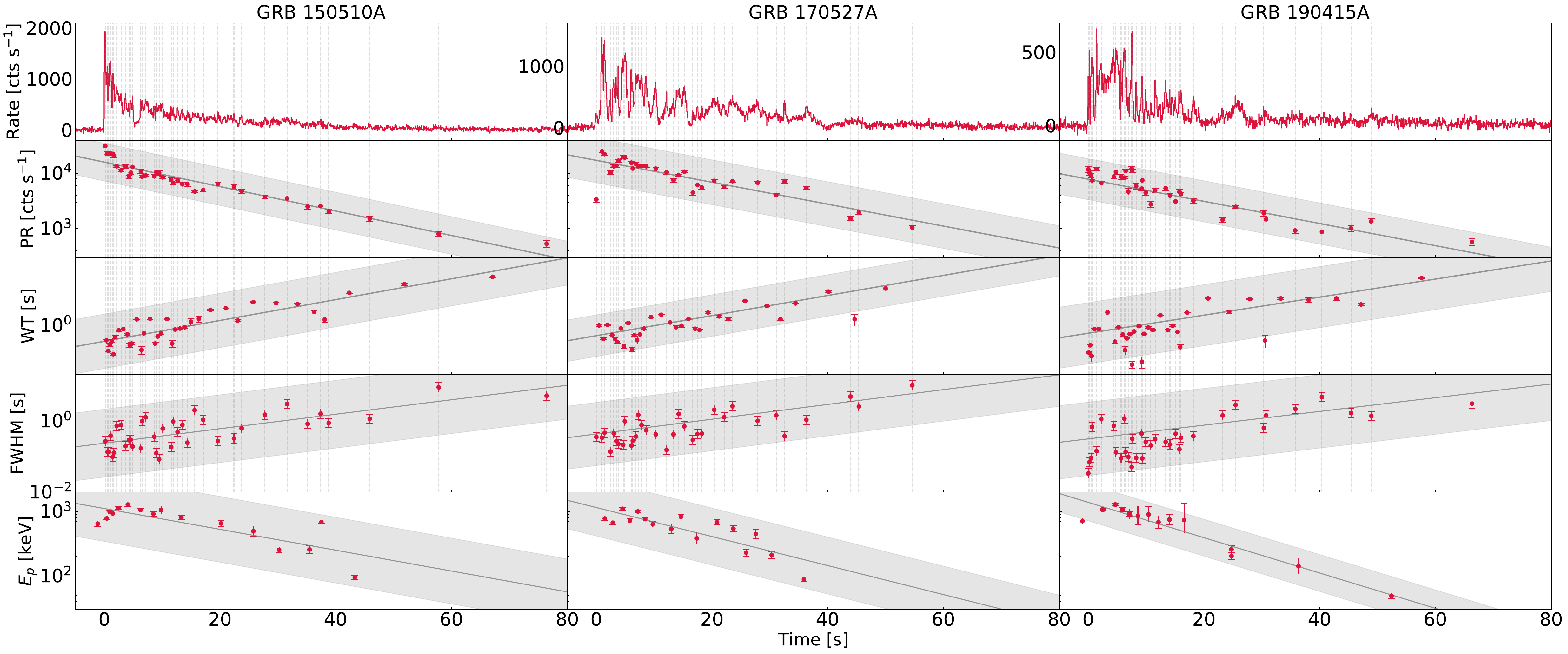}
\caption{Temporal evolution of the PR, WT, FWHM, and $E_{\mathrm{p}}$ for three GRBs selected from the sample $S_0$. The grey solid lines represent the best-fitting exponential evolution and the shaded areas show the 3$\sigma$ confidence intervals.}
\label{fig:evolution_three}
\end{figure*}
%
\subsection{Time-resolved spectral analysis}
\label{sec:time_res_spec_analysis}
Using the \GBM~data tools, we modelled the time-resolved energy spectra adopting the Band function \citep{Band93} and derived the temporal evolution of the peak energy $E_p$ of the $\nu\,F_\nu$ spectrum throughout the burst. The high-energy power-law index $\beta$ was fixed to $\beta = -2.3$ whenever it could not be constrained  (e.g., \citealt{KWGRBcat17}).
Time-resolved intervals were determined via Bayesian blocks \citep{Scargle13}, merging adjacent blocks to ensure $\geq 1000$ counts per interval \citep{Maistrello24}. The criterion on the minimum number of counts was added to ensure that each block provides a statistically meaningful fit of the spectrum.

For at least seven GRBs, we observed an exponential evolution of $E_p$ with time, sometimes following an initial rise before the peak. For GRB\,140306A and GRB\,230304B, we observed a significant tracking throughout the burst, in addition to the hard-to-soft exponential behaviour. For two cases, GRB\, 090831 and GRB\,221121A, we were unable to obtain the time-integrated spectrum, owing to the faintness of the signal. The results of the modelling are reported in Table~\ref{tab:Ep_fit_results} and the evolution of $E_{\rm p}$ with time is shown in Figs.~\ref{fig:evolution_three} and ~\ref{fig:evolution_six_others}.

\subsection{Study in the $E_{\rm p,i}-E_{\rm iso}$ plane}
\label{sec:EpiEiso}
We retrieved the intrinsic spectral peak energy $E_{\rm p,i} = E_{\rm p}\,(1+z)$ and isotropic-equivalent energy $E_{\rm iso}$ values for 344 type-II GRBs with known redshift $z$, using all available data from February 1997 to September 2025, mainly from the \textit{Konus}/Wind catalogues \citep{KWGRBcat17,KWGRBcat21}. Hereafter, this sample is referred to as $S_{\rm II}$. 

We modelled the $E_{\rm p,i}-E_{\rm iso}$ correlation with a power-law following standard procedures, using the D'Agostini likelihood \citep{DAgostini05}, which accounts for errors on both $x$- and $y$-axis and models the intrinsic dispersion $\sigma_{\rm int}$ of the relation. The likelihood maximisation was carried out within a Bayesian framework, performing a Markov Chain Monte Carlo analysis (MCMC), using the python package {\tt  emcee} \citep{Foreman13}. We modelled the relation as
\begin{equation}
\label{eq:amati_relation}
    \log \Big ( \frac{E_{\rm p,i}}{\rm keV} \Big ) = m \log\Big(\frac{E_{\rm iso}}{\rm erg}\Big) + q\;,
\end{equation}
obtaining $q=-15.3\pm 1.5$, $m=0.34\pm0.03$ and $\sigma_{\rm int} = 0.27\pm0.02$.
In parallel, we considered a sample of 47 type-I GRBs (15 of which being classified as short with extended emission), which is hereafter referred to as $S_{\rm I}$. Data were taken from the different \Konus catalogues \citep{Svinkin16,KWGRBcat17,KWGRBcat21,Lysenko25} and complemented with the data from the \GBM~ online catalogue\footnote{\url{ https://heasarc.gsfc.nasa.gov/w3browse/fermi/fermigbrst.html}}.

\begin{figure}
    \centering
    \includegraphics[width=0.45\textwidth]{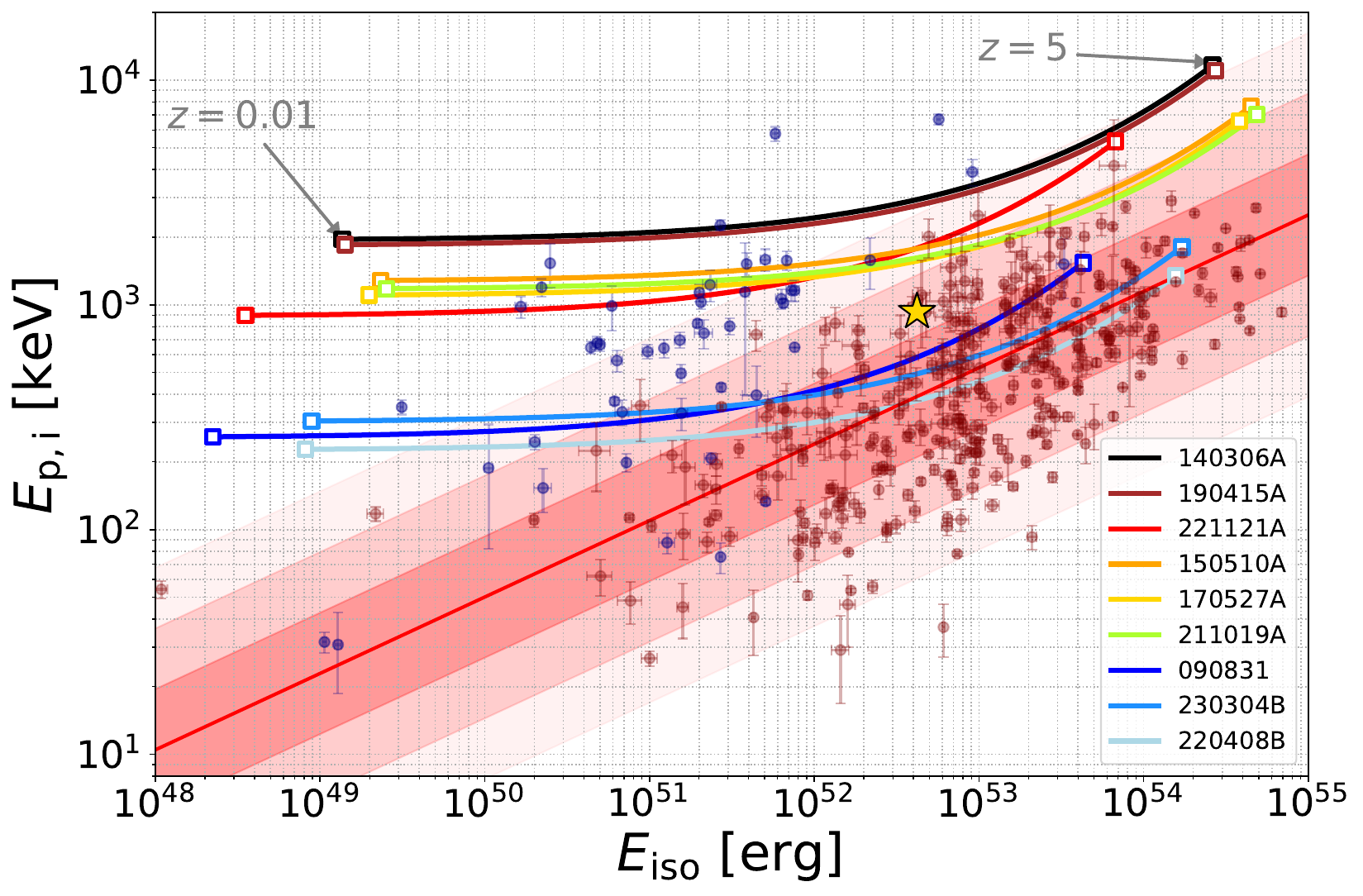}%
    \caption{Tracks of type-I candidates (sample $S_0$) in the $E_{\text{p,i}}$–$E_{\text{iso}}$ plane as a function of $z$. The  tracks are colour-coded according to their $\mathcal{L}_{*}$ values, with darker (lighter) colours corresponding to higher (lower) $\mathcal{L}_{*}$.
    Blue and brown points represent GRBs from sample $S_{\rm I}$ and $S_{\rm II}$, respectively.
    The red line shows the $E_{\text{p,i}}$–$E_{\text{iso}}$ relation and the shaded areas are the 1-, 2-, and 3-$\sigma_{\rm int}$ regions. The gold star is GRB\,230307A.}
    \label{fig:amati_tracks}
\end{figure}

All $S_0$ candidates have unknown redshifts. One can track their paths in the $E_{\rm p,i}-E_{\rm iso}$ plane as a function of the unknown redshift $z$ treated as a variable in the $0.01$--$5$ range. To this aim, we used the fluence $f$ and the observed peak energy $E_{\rm p}$ from \GBM~catalogue. The isotropic-equivalent energy $E_{\rm iso}(z)$ was calculated as $E_{\rm iso}(z)\,=\,4 \pi d^2_L(z)f/(1+z)$,
where $d_L(z)$ is the luminosity distance, while the intrinsic peak energy is $E_{\rm p,i}(z) = E_{\rm p}\,(1+z)$. Figure~\ref{fig:amati_tracks} shows the results. Overall, one cannot help but notice that the $S_0$ candidates preferentially populate the region above the $E_{\rm p,i}$-$E_{\rm iso}$ relation, where most type-I GRBs lie.

To evaluate the joint probability that the location of the $S_0$ sample in the $E_{\rm iso}$-$E_{\rm p,i}$ plane is incidental and that they are type-II GRBs, compatibly with the relation dispersion $\sigma_{\rm int}$ and with the individual measurement uncertainties, we conceived the following test.
We collected all long ($T_{90}>2$~s) GRBs without redshift in the online \GBM~ catalogue (excluding the $S_0$ candidates) into the sample called $S_{\rm IInz}$. Under the assumption that all these GRBs are consistent with the $E_{\rm p,i}$-$E_{\rm iso}$ relation, for each of them we derived the path in the $E_{\rm iso}$--$E_{\rm p,i}$ plane and calculated the D'Agostini negative log-likelihood that best models the $E_{\rm p,i}$-$E_{\rm iso}$ of $S_{\rm II}$, whose parameters $(m, q, \sigma_{\rm int})$ are reported above:
\begin{equation}
\label{eq:likelihood}
   \mathcal{L}(z) = \frac{1}{2}\log\Big [2\pi (\sigma_{\rm int}^{2}+\sigma_{y}^{2} + m^{2}\,\sigma_{x}^{2})\Big] + \frac{[y(z) -m\,x(z)-q]^{2}}{2(\sigma_{\rm int}^2+\sigma_{y}^{2}+m^2\,\sigma_{x}^2)},
\end{equation}
where $x(z) = \log E_{\rm iso}(z)$ and $y(z) = \log E_{\rm p,i}(z)$ and $\sigma_x$ and $\sigma_{y}$ are the corresponding uncertainties, which do not depend on $z$, but just on the relative uncertainties on $f$ and on $E_{\rm p}$.

For each GRB of $S_{\rm IInz}$ we determined $z_{*}$, that is the value of $z$ minimising $\mathcal{L}(z)$: $\mathcal{L}_{*} = \mathcal{L}(z_{*}) = \min_z{(\mathcal{L}(z))}$.
The same procedure was applied to the $S_0$ sample, whose $\mathcal{L}_*$ values are reported in Table~\ref{tab:grb_type_I_candidates}. 
The sum of the values obtained for the $S_0$ sample is a measure of the un-normalised joint likelihood, being the logarithm of the product of all individual likelihood values: $\mathcal{L}_{*,S_0}^{\rm (tot)} = \sum_{i=1}^{9} \mathcal{L}_{*}^{(i)}$, where $\mathcal{L}_{*}^{(i)}$ is the value for the $i$-th GRB of $S_0$.

To determine the $p$-value, that is the probability of obtaining a value $\geq$ $\mathcal{L}_{*,S_0}^{\rm (tot)}$ under the assumption that they are all type-II GRBs, we built the reference distribution for $\mathcal{L}_{*,S_0}^{\rm (tot)}$ as follows.
We randomly drew $N_{\rm sim} = 10^6$ samples of $N=9$ GRBs from $S_{\rm IInz}$ and for each of them we calculated the same metric, that is $\mathcal{L}_{*}^{\rm (tot)}.$ The corresponding distribution of $\mathcal{L}_{*}^{\rm (tot)}$ values was then used as a reference. As a result, only 124 cases had $\mathcal{L}_{*}^{\rm (tot)}\ge \mathcal{L}_{*,S_0}^{\rm (tot)}.$ We conclude that the assumption that our $S_0$ sample of 9 long merger candidates are all type-II GRBs following the $E_{\rm p,i}-E_{\rm iso}$ relation has a $p$-value equal to $1.24 \times 10^{-4}$, equivalent to $3.7\,\sigma$ (Gaussian). 
We also tried to reject the following more general $H_0$ hypothesis: $S_0$ includes at most $N_{\rm out}$ outliers of the $E_{\rm p,i}-E_{\rm iso}$ relation. To test this $H_0$, one has to reject from each sample a number $N_{\rm out}$ of GRBs with the largest $\mathcal{L}_*$ values out of 9 and calculate the total value over the remaining ones, $\mathcal{L}_{*}^{\rm (tot)} = \sum_{i=1}^{9-N_{\rm out}} \mathcal{L}_{*}^{(i)}$. This must be done for the real and the simulated samples. The case $N_{\rm out}=0$ is already done above. Table~\ref{tab:h0_tests} reports the $p$-values obtained for a range of $N_{\rm out}$ values. In particular, for $N_{\rm out}=1$ the $p$-value is $8.21\times10^{-4}$, equivalent to $3.1\,\sigma$ (Gaussian). Thus, with $>3\sigma$ confidence we conclude that at least 2/9 of our $S_0$ candidates do not satisfy the $E_{\rm p,i}-E_{\rm iso}$ relation.
%
\subsection{Analysis in the MVT-$T_{\rm 90}$ plane}
\label{sec:MVT_T90}
\begin{figure}
    \centering
    \includegraphics[width=0.45\textwidth]{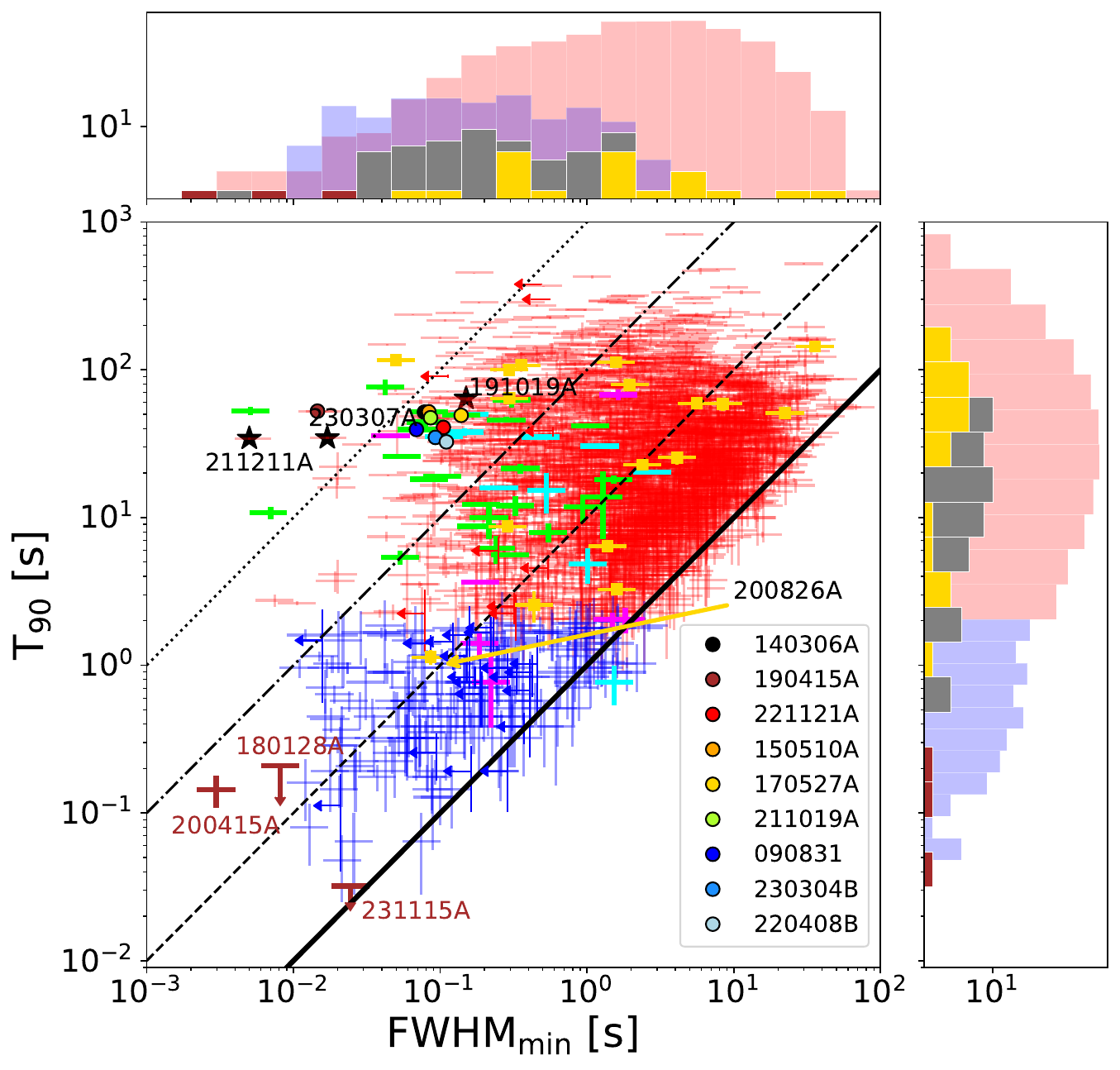}
    \caption{MVT-$T_{\rm 90}$ plane, with MVT defined as $\rm FWHM_{min}$ (adapted from \citealt{Maccary25}).
    The filled circles represent the type-I GRB candidates (sample $S_0$) identified in the present analysis. The blue and red crosses indicate short ($T_{90}<2~s$) and long ($T_{90}>2~s)$ GRBs, respectively. Gold points represent SN-associated GRBs. Magenta, lime, and cyan points represent SEE-GRBs from \citet{Lien16,Lan20,Kaneko15}, respectively. Three extragalactic magnetar giant flare candidates, 180128A, 200415A, and 231115A are shown in brown. The solid black line represents the equality line while the dashed, dashed-dotted, and dotted lines represent a factor of 10, 100, and 1000 respectively, deviation from equality.
    } 
    \label{fig:mvt_t90}
\end{figure}

Figure~\ref{fig:mvt_t90} shows the $S_0$ candidates in the MVT-$T_{\rm 90}$ plane. The data set of \citet{Maccary25} was used as a reference.
We adopt the MVT definition based on FWHM$_{\rm min}$, which is the FWHM of the shortest pulse, following \citet{Camisasca23}, although similar conclusions can be drawn using alternative definitions\footnote{$\rm FWHM_{\rm min}$ correlates with other MVT metrics \citep{Golkhou15,MacLachlan13}, with values that are on average $\sim 5-6$ larger \citep{Maccary25}. This difference does not affect the results presented here, since the relative position of the bursts in the MVT-$T_{\rm 90}$ plane remains essentially unchanged.}. In Appendix~\ref{sec:impact_MVT_metrics}, we studied the impact of using different MVT metrics.
Both quantities were not divided by $(1+z)$ for the reasons explained in \citet{Camisasca23}.
All $S_0$ candidates lie at similar locations in the upper left of the plot, with $T_{\rm 90}\sim 30-50~\rm s$ and MVT $\lesssim0.1$~s. Notably, some previously identified long type-I GRBs, such as GRB\,191019A, GRB\,211211A, and GRB\,230307A, lie in the same region.
In particular, GRB\,190415A and GRB\,230307A lie very close to each other. Only $0.2\%$ of long bursts exhibit smaller MVT values than GRB\,190415A. MVT and $T_{\rm 90}$ values of GRBs from the $S_0$ set are reported in Table~\ref{tab:grb_type_I_candidates}.
%

\section{Discussion and conclusion}
\label{sec:discussion}
We identified 9 type-I candidates sharing properties similar to those noted in \citetalias{Maccary26} and analysed their positions in the \(E_{\text{p,i}}-E_{\text{iso}}\) plane as a function of (unknown) redshift. Out of 263 bursts selected from the full \textit{Fermi}/GBM catalogue, the fraction of long, multi-peaked type-I candidates identified via this method is around 3\%. We demonstrated, with a confidence level exceeding $3 \sigma$, that at least two of these candidates are incompatible with the population of long GRBs lacking measured redshifts.

This suggests that the distinctive properties identified in \citetalias{Maccary26} serve as a promising indicator of a compact object merger (COM) origin. Such findings have practical applications for identifying COM-associated GRBs, as these characteristics can be extracted directly from the prompt-emission light curve. This provides early, direct clues of a COM origin, which can help guide and prioritise the search for multi-wavelength counterparts. Unfortunately, owing to the poor localisations of these events, no afterglow searches were conducted for any of these candidates. Consequently, their redshifts remain unknown. In Appendix \ref{sec:ul_redshfit}, based on the possible incompatibility of these bursts with the Amati relation, we propose some limiting values for their redshifts. A preliminary search for the best-localised candidate, GRB\,230304B\footnote{It was localised by the InterPlanetary Network within an error box of about $100$ arcmin square (\citealt{Svinkin23_230304B}).}, identifies potential hosts at $z\lesssim 0.04$. This would render the event consistent with other known long mergers in terms of energetics and luminosity, although the precise identification of the host remains unfeasible at this stage.
Prospectively, promptly identifying these distinctive properties in future GRBs could help the community to quickly select promising COM candidates and plan dedicated multi-wavelength campaigns aimed at detecting the afterglow, assessing and characterising the progenitor.
These distinctive properties are not only observationally significant, but also challenge the stochastic nature predicted by most theoretical models of GRB prompt emission. Specifically, the observed increase in pulse width throughout the burst profile is difficult to reconcile with the internal shock model; previously, the lack of such evolution in various multi-peaked GRBs was used to favour internal shocks over external ones (e.g., \citealt{Fenimore99}). However, this reasoning applies neither to GRB\,230307A nor to the other GRBs identified in this study.
The observation of these properties thus opens the door for alternative scenarios, such as the target-shell model introduced in \citetalias{Maccary26} – which posits the existence of a primitive shell emitted before the main train of shells from the central engine
– or within refined versions of the ICMART framework \citep{ICMART}, where an expanding emission region with mini-emitters radiating through magnetic reconnection accounts for the properties of long merger events like GRB\,230307A \citep{Yi25}.

\begin{acknowledgements}
A. Rossi acknowledges financial support from INAF Mini Grant RSN4 (ID: 1.05.24.07.04).
\end{acknowledgements}

\bibliographystyle{aa}
\bibliography{alles_grbs,alles_grbs2}

@ARTICLE{Amati02,
   author = {{Amati}, L. and {Frontera}, F. and {Tavani}, M. and {in't Zand}, J.~J.~M. and 
	{Antonelli}, A. and {Costa}, E. and {Feroci}, M. and {Guidorzi}, C. and 
	{Heise}, J. and {Masetti}, N. and {Montanari}, E. and {Nicastro}, L. and 
	{Palazzi}, E. and {Pian}, E. and {Piro}, L. and {Soffitta}, P.},
    title = "{Intrinsic spectra and energetics of BeppoSAX Gamma-Ray Bursts with known redshifts}",
  journal = {A\&A},
   eprint = {arXiv:astro-ph/0205230},
     year = 2002,
    month = jul,
   volume = 390,
    pages = {81-89},
      doi = {10.1051/0004-6361:20020722},
   adsurl = {http://adsabs.harvard.edu/abs/2002A%26A...390...81A}
}

@ARTICLE{Amati06,
   author = {{Amati}, L.},
    title = "{The E$_{p,i}$-E$_{iso}$ correlation in gamma-ray bursts: updated observational status, re-analysis and main implications}",
  journal = {MNRAS},
   eprint = {arXiv:astro-ph/0601553},
     year = 2006,
    month = oct,
   volume = 372,
    pages = {233-245},
      doi = {10.1111/j.1365-2966.2006.10840.x},
   adsurl = {http://adsabs.harvard.edu/abs/2006MNRAS.372..233A}
}

@ARTICLE{Band93,
   author = {{Band}, D. and {Matteson}, J. and {Ford}, L. and {Schaefer}, B. and 
	{Palmer}, D. and {Teegarden}, B. and {Cline}, T. and {Briggs}, M. and 
	{Paciesas}, W. and {Pendleton}, G. and {Fishman}, G. and {Kouveliotou}, C. and 
	{Meegan}, C. and {Wilson}, R. and {Lestrade}, P.},
    title = "{BATSE observations of gamma-ray burst spectra. I - Spectral diversity}",
  journal = {ApJ},
     year = 1993,
    month = aug,
   volume = 413,
    pages = {281-292},
      doi = {10.1086/172995},
   adsurl = {http://adsabs.harvard.edu/abs/1993ApJ...413..281B}
}

@ARTICLE{Camisasca23,
       author = {{Camisasca}, A.~E. and {Guidorzi}, C. and {Amati}, L. and {Frontera}, F. and {Song}, X.~Y. and {Xiao}, S. and {Xiong}, S.~L. and {Zhang}, S.~N. and {Margutti}, R. and {Kobayashi}, S. and {Mundell}, C.~G. and {Ge}, M.~Y. and {Gomboc}, A. and {Jia}, S.~M. and {Jordana-Mitjans}, N. and {Li}, C.~K. and {Li}, X.~B. and {Maccary}, R. and {Shrestha}, M. and {Xue}, W.~C. and {Zhang}, S.},
        title = "{GRB minimum variability timescale with Insight-HXMT and Swift. Implications for progenitor models, dissipation physics, and GRB classifications}",
      journal = {\aap},
         year = 2023,
        month = mar,
       volume = {671},
          eid = {A112},
        pages = {A112},
          doi = {10.1051/0004-6361/202245657},
archivePrefix = {arXiv},
       eprint = {2301.01176},
 primaryClass = {astro-ph.HE},
       adsurl = {https://ui.adsabs.harvard.edu/abs/2023A&A...671A.112C}
}

@ARTICLE{cosmoPlanck20,
       author = {{Planck Collaboration} and {Aghanim}, N. and {Akrami}, Y. and {Ashdown}, M. and {Aumont}, J. and {Baccigalupi}, C. and {Ballardini}, M. and {Banday}, A.~J. and {Barreiro}, R.~B. and {Bartolo}, N. and {Basak}, S. and {Battye}, R. and {Benabed}, K. and {Bernard}, J. -P. and {Bersanelli}, M. and {Bielewicz}, P. and {Bock}, J.~J. and {Bond}, J.~R. and {Borrill}, J. and {Bouchet}, F.~R. and {Boulanger}, F. and {Bucher}, M. and {Burigana}, C. and {Butler}, R.~C. and {Calabrese}, E. and {Cardoso}, J. -F. and {Carron}, J. and {Challinor}, A. and {Chiang}, H.~C. and {Chluba}, J. and {Colombo}, L.~P.~L. and {Combet}, C. and {Contreras}, D. and {Crill}, B.~P. and {Cuttaia}, F. and {de Bernardis}, P. and {de Zotti}, G. and {Delabrouille}, J. and {Delouis}, J. -M. and {Di Valentino}, E. and {Diego}, J.~M. and {Dor{\'e}}, O. and {Douspis}, M. and {Ducout}, A. and {Dupac}, X. and {Dusini}, S. and {Efstathiou}, G. and {Elsner}, F. and {En{\ss}lin}, T.~A. and {Eriksen}, H.~K. and {Fantaye}, Y. and {Farhang}, M. and {Fergusson}, J. and {Fernandez-Cobos}, R. and {Finelli}, F. and {Forastieri}, F. and {Frailis}, M. and {Fraisse}, A.~A. and {Franceschi}, E. and {Frolov}, A. and {Galeotta}, S. and {Galli}, S. and {Ganga}, K. and {G{\'e}nova-Santos}, R.~T. and {Gerbino}, M. and {Ghosh}, T. and {Gonz{\'a}lez-Nuevo}, J. and {G{\'o}rski}, K.~M. and {Gratton}, S. and {Gruppuso}, A. and {Gudmundsson}, J.~E. and {Hamann}, J. and {Handley}, W. and {Hansen}, F.~K. and {Herranz}, D. and {Hildebrandt}, S.~R. and {Hivon}, E. and {Huang}, Z. and {Jaffe}, A.~H. and {Jones}, W.~C. and {Karakci}, A. and {Keih{\"a}nen}, E. and {Keskitalo}, R. and {Kiiveri}, K. and {Kim}, J. and {Kisner}, T.~S. and {Knox}, L. and {Krachmalnicoff}, N. and {Kunz}, M. and {Kurki-Suonio}, H. and {Lagache}, G. and {Lamarre}, J. -M. and {Lasenby}, A. and {Lattanzi}, M. and {Lawrence}, C.~R. and {Le Jeune}, M. and {Lemos}, P. and {Lesgourgues}, J. and {Levrier}, F. and {Lewis}, A. and {Liguori}, M. and {Lilje}, P.~B. and {Lilley}, M. and {Lindholm}, V. and {L{\'o}pez-Caniego}, M. and {Lubin}, P.~M. and {Ma}, Y. -Z. and {Mac{\'\i}as-P{\'e}rez}, J.~F. and {Maggio}, G. and {Maino}, D. and {Mandolesi}, N. and {Mangilli}, A. and {Marcos-Caballero}, A. and {Maris}, M. and {Martin}, P.~G. and {Martinelli}, M. and {Mart{\'\i}nez-Gonz{\'a}lez}, E. and {Matarrese}, S. and {Mauri}, N. and {McEwen}, J.~D. and {Meinhold}, P.~R. and {Melchiorri}, A. and {Mennella}, A. and {Migliaccio}, M. and {Millea}, M. and {Mitra}, S. and {Miville-Desch{\^e}nes}, M. -A. and {Molinari}, D. and {Montier}, L. and {Morgante}, G. and {Moss}, A. and {Natoli}, P. and {N{\o}rgaard-Nielsen}, H.~U. and {Pagano}, L. and {Paoletti}, D. and {Partridge}, B. and {Patanchon}, G. and {Peiris}, H.~V. and {Perrotta}, F. and {Pettorino}, V. and {Piacentini}, F. and {Polastri}, L. and {Polenta}, G. and {Puget}, J. -L. and {Rachen}, J.~P. and {Reinecke}, M. and {Remazeilles}, M. and {Renzi}, A. and {Rocha}, G. and {Rosset}, C. and {Roudier}, G. and {Rubi{\~n}o-Mart{\'\i}n}, J.~A. and {Ruiz-Granados}, B. and {Salvati}, L. and {Sandri}, M. and {Savelainen}, M. and {Scott}, D. and {Shellard}, E.~P.~S. and {Sirignano}, C. and {Sirri}, G. and {Spencer}, L.~D. and {Sunyaev}, R. and {Suur-Uski}, A. -S. and {Tauber}, J.~A. and {Tavagnacco}, D. and {Tenti}, M. and {Toffolatti}, L. and {Tomasi}, M. and {Trombetti}, T. and {Valenziano}, L. and {Valiviita}, J. and {Van Tent}, B. and {Vibert}, L. and {Vielva}, P. and {Villa}, F. and {Vittorio}, N. and {Wandelt}, B.~D. and {Wehus}, I.~K. and {White}, M. and {White}, S.~D.~M. and {Zacchei}, A. and {Zonca}, A.},
        title = "{Planck 2018 results. VI. Cosmological parameters}",
      journal = {\aap},
         year = 2020,
        month = sep,
       volume = {641},
          eid = {A6},
        pages = {A6},
          doi = {10.1051/0004-6361/201833910},
archivePrefix = {arXiv},
       eprint = {1807.06209},
 primaryClass = {astro-ph.CO},
       adsurl = {https://ui.adsabs.harvard.edu/abs/2020A&A...641A...6P}
}

@ARTICLE{DAgostini05,
   author = {{D'Agostini}, G.},
    title = "{Fits, and especially linear fits, with errors on both axes, extra variance of the data points and other complications}",
  journal = {ArXiv Physics e-prints},
   eprint = {physics/0511182},
     year = 2005,
    month = nov,
   adsurl = {http://adsabs.harvard.edu/abs/2005physics..11182D}
}

@ARTICLE{Dai24,
       author = {{Dai}, Cui-Yuan and {Guo}, Chen-Lei and {Zhang}, Hai-Ming and {Liu}, Ruo-Yu and {Wang}, Xiang-Yu},
        title = "{Evidence for a Compact Stellar Merger Origin for GRB 230307A From Fermi-LAT and Multiwavelength Afterglow Observations}",
      journal = {\apjl},
         year = 2024,
        month = feb,
       volume = {962},
       number = {2},
          eid = {L37},
        pages = {L37},
          doi = {10.3847/2041-8213/ad2680},
archivePrefix = {arXiv},
       eprint = {2312.01074},
 primaryClass = {astro-ph.HE},
       adsurl = {https://ui.adsabs.harvard.edu/abs/2024ApJ...962L..37D}
}

@ARTICLE{Fenimore99,
   author = {{Fenimore}, E.~E. and {Ramirez-Ruiz}, E. and {Wu}, B.},
    title = "{GRB 990123: Evidence that the Gamma Rays Come from a Central Engine}",
  journal = {ApJ},
   eprint = {astro-ph/9902007},
     year = 1999,
    month = jun,
   volume = 518,
    pages = {L73-L76},
      doi = {10.1086/312075},
   adsurl = {http://adsabs.harvard.edu/abs/1999ApJ...518L..73F}
}

@ARTICLE{Foreman13,
       author = {{Foreman-Mackey}, Daniel and {Hogg}, David W. and {Lang}, Dustin and
         {Goodman}, Jonathan},
        title = "{emcee: The MCMC Hammer}",
      journal = {\pasp},
         year = "2013",
        month = "Mar",
       volume = {125},
       number = {925},
        pages = {306},
          doi = {10.1086/670067},
archivePrefix = {arXiv},
       eprint = {1202.3665},
 primaryClass = {astro-ph.IM},
       adsurl = {https://ui.adsabs.harvard.edu/abs/2013PASP..125..306F}
}

@ARTICLE{Gillanders25,
       author = {{Gillanders}, J.~H. and {Smartt}, S.~J.},
        title = "{Analysis of the JWST spectra of the kilonova AT 2023vfi accompanying GRB 230307A}",
      journal = {\mnras},
         year = 2025,
        month = apr,
       volume = {538},
       number = {3},
        pages = {1663-1689},
          doi = {10.1093/mnras/staf287},
archivePrefix = {arXiv},
       eprint = {2408.11093},
 primaryClass = {astro-ph.HE},
       adsurl = {https://ui.adsabs.harvard.edu/abs/2025MNRAS.538.1663G}
}

@ARTICLE{Golkhou14,
   author = {{Golkhou}, V.~Z. and {Butler}, N.~R.},
    title = "{Uncovering the Intrinsic Variability of Gamma-Ray Bursts}",
  journal = {ApJ},
archivePrefix = "arXiv",
   eprint = {1403.4254},
 primaryClass = "astro-ph.HE",
     year = 2014,
    month = may,
   volume = 787,
      eid = {90},
    pages = {90},
      doi = {10.1088/0004-637X/787/1/90},
   adsurl = {http://adsabs.harvard.edu/abs/2014ApJ...787...90G}
}

@ARTICLE{Golkhou15,
   author = {{Golkhou}, V.~Z. and {Butler}, N.~R. and {Littlejohns}, O.~M.},
    title = "{The Energy Dependence of GRB Minimum Variability Timescales}",
  journal = {ApJ},
archivePrefix = "arXiv",
   eprint = {1501.05948},
 primaryClass = "astro-ph.HE",
     year = 2015,
    month = oct,
   volume = 811,
      eid = {93},
    pages = {93},
      doi = {10.1088/0004-637X/811/2/93},
   adsurl = {http://adsabs.harvard.edu/abs/2015ApJ...811...93G}
}

@article{Guidorzi15a,
  author = {{Guidorzi}, C.},
  title = "{MEPSA: a flexible peak search algorithm designed for uniformly spaced time series}",
  journal = {Astronomy and Computing},
  volume = 10,
  pages = {54-60},
  month = apr,
  year = 2015,
  issn = {2213-1337},
  doi = {10.1016/j.ascom.2015.01.001},
  url = {http://www.sciencedirect.com/science/article/pii/S2213133715000025}
}

@ARTICLE{Kaneko15,
       author = {{Kaneko}, Y. and {Bostanc{\i}}, Z.~F. and {G{\"o}{\u{g}}{\"u}{\c{s}}}, E. and {Lin}, L.},
        title = "{Short gamma-ray bursts with extended emission observed with Swift/BAT and Fermi/GBM}",
      journal = {\mnras},
         year = 2015,
        month = sep,
       volume = {452},
       number = {1},
        pages = {824-837},
          doi = {10.1093/mnras/stv1286},
archivePrefix = {arXiv},
       eprint = {1506.05899},
 primaryClass = {astro-ph.HE},
       adsurl = {https://ui.adsabs.harvard.edu/abs/2015MNRAS.452..824K}
}

@ARTICLE{KWGRBcat17,
   author = {{Tsvetkova}, A. and {Frederiks}, D. and {Golenetskii}, S. and {Lysenko}, A. and {Oleynik}, P. and {Pal'shin}, V. and {Svinkin}, D. and {Ulanov}, M. and {Cline}, T. and {Hurley}, K. and {Aptekar}, R.},
    title = "{The Konus-Wind Catalog of Gamma-Ray Bursts with Known Redshifts. I. Bursts Detected in the Triggered Mode}",
  journal = {\apj},
archivePrefix = "arXiv",
   eprint = {1710.08746},
 primaryClass = "astro-ph.HE",
     year = 2017,
    month = dec,
   volume = 850,
      eid = {161},
    pages = {161},
      doi = {10.3847/1538-4357/aa96af},
   adsurl = {http://adsabs.harvard.edu/abs/2017ApJ...850..161T}
}

@ARTICLE{KWGRBcat21,
       author = {{Tsvetkova}, Anastasia and {Frederiks}, Dmitry and {Svinkin}, Dmitry and {Aptekar}, Rafail and {Cline}, Thomas L. and {Golenetskii}, Sergei and {Hurley}, Kevin and {Lysenko}, Alexandra and {Ridnaia}, Anna and {Ulanov}, Mikhail},
        title = "{The Konus-Wind Catalog of Gamma-Ray Bursts with Known Redshifts. II. Waiting-Mode Bursts Simultaneously Detected by Swift/BAT}",
      journal = {\apj},
         year = 2021,
        month = feb,
       volume = {908},
       number = {1},
          eid = {83},
        pages = {83},
          doi = {10.3847/1538-4357/abd569},
archivePrefix = {arXiv},
       eprint = {2012.14849},
 primaryClass = {astro-ph.HE},
       adsurl = {https://ui.adsabs.harvard.edu/abs/2021ApJ...908...83T}
}

@ARTICLE{Lan20,
       author = {{Lan}, Lin and {Lu}, Rui-Jingi and {L{\"u}}, Hou-Jun and {Shen}, Jun and {Rice}, Jared and {Li}, Long and {Liang}, En-Wei},
        title = "{The properties of prompt emission in short gamma-ray bursts with extended emission observed by Fermi/GBM}",
      journal = {\mnras},
         year = 2020,
        month = mar,
       volume = {492},
       number = {3},
        pages = {3622-3630},
          doi = {10.1093/mnras/staa044},
archivePrefix = {arXiv},
       eprint = {2001.01150},
 primaryClass = {astro-ph.HE},
       adsurl = {https://ui.adsabs.harvard.edu/abs/2020MNRAS.492.3622L}
}

@article{Levan24,
       author = {{Levan}, Andrew J. and {Gompertz}, Benjamin P. and {Salafia}, Om Sharan and {Bulla}, Mattia and {Burns}, Eric and {Hotokezaka}, Kenta and {Izzo}, Luca and {Lamb}, Gavin P. and {Malesani}, Daniele B. and {Oates}, Samantha R. and {Ravasio}, Maria Edvige and {Rouco Escorial}, Alicia and {Schneider}, Benjamin and {Sarin}, Nikhil and {Schulze}, Steve and {Tanvir}, Nial R. and {Ackley}, Kendall and {Anderson}, Gemma and {Brammer}, Gabriel B. and {Christensen}, Lise and {Dhillon}, Vikram S. and {Evans}, Phil A. and {Fausnaugh}, Michael and {Fong}, Wen-fai and {Fruchter}, Andrew S. and {Fryer}, Chris and {Fynbo}, Johan P.~U. and {Gaspari}, Nicola and {Heintz}, Kasper E. and {Hjorth}, Jens and {Kennea}, Jamie A. and {Kennedy}, Mark R. and {Laskar}, Tanmoy and {Leloudas}, Giorgos and {Mandel}, Ilya and {Martin-Carrillo}, Antonio and {Metzger}, Brian D. and {Nicholl}, Matt and {Nugent}, Anya and {Palmerio}, Jesse T. and {Pugliese}, Giovanna and {Rastinejad}, Jillian and {Rhodes}, Lauren and {Rossi}, Andrea and {Saccardi}, Andrea and {Smartt}, Stephen J. and {Stevance}, Heloise F. and {Tohuvavohu}, Aaron and {van der Horst}, Alexander and {Vergani}, Susanna D. and {Watson}, Darach and {Barclay}, Thomas and {Bhirombhakdi}, Kornpob and {Breedt}, Elm{\'e} and {Breeveld}, Alice A. and {Brown}, Alexander J. and {Campana}, Sergio and {Chrimes}, Ashley A. and {D'Avanzo}, Paolo and {D'Elia}, Valerio and {De Pasquale}, Massimiliano and {Dyer}, Martin J. and {Galloway}, Duncan K. and {Garbutt}, James A. and {Green}, Matthew J. and {Hartmann}, Dieter H. and {Jakobsson}, P{\'a}ll and {Kerry}, Paul and {Kouveliotou}, Chryssa and {Langeroodi}, Danial and {Le Floc'h}, Emeric and {Leung}, James K. and {Littlefair}, Stuart P. and {Munday}, James and {O'Brien}, Paul and {Parsons}, Steven G. and {Pelisoli}, Ingrid and {Sahman}, David I. and {Salvaterra}, Ruben and {Sbarufatti}, Boris and {Steeghs}, Danny and {Tagliaferri}, Gianpiero and {Th{\"o}ne}, Christina C. and {de Ugarte Postigo}, Antonio and {Kann}, David Alexander},
        title = "{Heavy-element production in a compact object merger observed by JWST}",
      journal = {\nat},
         year = 2024,
        month = feb,
       volume = {626},
       number = {8000},
        pages = {737-741},
          doi = {10.1038/s41586-023-06759-1},
archivePrefix = {arXiv},
       eprint = {2307.02098},
 primaryClass = {astro-ph.HE},
       adsurl = {https://ui.adsabs.harvard.edu/abs/2024Natur.626..737L}
}

@ARTICLE{Lien16,
   author = {{Lien}, A. and {Sakamoto}, T. and {Barthelmy}, S.~D. and {Baumgartner}, W.~H. and 
	{Cannizzo}, J.~K. and {Chen}, K. and {Collins}, N.~R. and {Cummings}, J.~R. and 
	{Gehrels}, N. and {Krimm}, H.~A. and {Markwardt}, C.~B. and 
	{Palmer}, D.~M. and {Stamatikos}, M. and {Troja}, E. and {Ukwatta}, T.~N.},
    title = "{The Third Swift Burst Alert Telescope Gamma-Ray Burst Catalog}",
  journal = {ApJ},
archivePrefix = "arXiv",
   eprint = {1606.01956},
 primaryClass = "astro-ph.HE",
     year = 2016,
    month = sep,
   volume = 829,
      eid = {7},
    pages = {7},
      doi = {10.3847/0004-637X/829/1/7},
   adsurl = {http://cdsads.u-strasbg.fr/abs/2016ApJ...829....7L}
}

@ARTICLE{Maccary25,
       author = {{Maccary}, R. and {Guidorzi}, C. and {Camisasca}, A.~E. and {Maistrello}, M. and {Kobayashi}, S. and {Amati}, L. and {Bazzanini}, L. and {Bulla}, M. and {Ferro}, L. and {Frontera}, F. and {Tsvetkova}, A.},
        title = "{Gamma-ray burst minimum variability timescales with Fermi/GBM}",
      journal = {\aap},
         year = 2025,
        month = oct,
       volume = {702},
          eid = {A95},
        pages = {A95},
          doi = {10.1051/0004-6361/202555418},
archivePrefix = {arXiv},
       eprint = {2508.08995},
 primaryClass = {astro-ph.HE},
       adsurl = {https://ui.adsabs.harvard.edu/abs/2025A&A...702A..95M}
}

@ARTICLE{Maccary26,
       author = {{Maccary}, R. and {Guidorzi}, C. and {Maistrello}, M. and {Kobayashi}, S. and {Bulla}, M. and {Moradi}, R. and {Yi}, S. -X. and {Wang}, C.~W. and {Zhang}, W.~L. and {Tan}, W. -J. and {Xiong}, S. -L. and {Zhang}, S. -N.},
        title = "{A set of distinctive properties ruling the prompt emission of GRB 230307A and other long {\ensuremath{\gamma}}-ray bursts from compact object mergers}",
      journal = {Journal of High Energy Astrophysics},
         year = 2026,
        month = jan,
       volume = {49},
          eid = {100456},
        pages = {100456},
          doi = {10.1016/j.jheap.2025.100456},
archivePrefix = {arXiv},
       eprint = {2509.05628},
 primaryClass = {astro-ph.HE},
       adsurl = {https://ui.adsabs.harvard.edu/abs/2026JHEAp..4900456M}
}

@ARTICLE{MacLachlan12,
       author = {{MacLachlan}, G.~A. and {Shenoy}, A. and {Sonbas}, E. and {Dhuga}, K.~S. and {Eskandarian}, A. and {Maximon}, L.~C. and {Parke}, W.~C.},
        title = "{The minimum variability time-scale and its relation to pulse profiles of Fermi GRBs}",
      journal = {\mnras},
         year = 2012,
        month = sep,
       volume = {425},
       number = {1},
        pages = {L32-L35},
          doi = {10.1111/j.1745-3933.2012.01295.x},
archivePrefix = {arXiv},
       eprint = {1205.0055},
 primaryClass = {astro-ph.HE},
       adsurl = {https://ui.adsabs.harvard.edu/abs/2012MNRAS.425L..32M}
}

@ARTICLE{MacLachlan13,
       author = {{MacLachlan}, G.~A. and {Shenoy}, A. and {Sonbas}, E. and {Dhuga}, K.~S. and {Cobb}, B.~E. and {Ukwatta}, T.~N. and {Morris}, D.~C. and {Eskandarian}, A. and {Maximon}, L.~C. and {Parke}, W.~C.},
        title = "{Minimum variability time-scales of long and short GRBs}",
      journal = {\mnras},
         year = 2013,
        month = jun,
       volume = {432},
       number = {2},
        pages = {857-865},
          doi = {10.1093/mnras/stt241},
archivePrefix = {arXiv},
       eprint = {1201.4431},
 primaryClass = {astro-ph.HE},
       adsurl = {https://ui.adsabs.harvard.edu/abs/2013MNRAS.432..857M}
}

@ARTICLE{Maistrello24,
       author = {{Maistrello}, M. and {Maccary}, R. and {Guidorzi}, C. and {Amati}, L.},
        title = "{The dispersion of the E$_{p, i}$-L$_{iso}$ correlation of long gamma-ray bursts is partially due to assembling different sources}",
      journal = {\aap},
         year = 2024,
        month = apr,
       volume = {684},
          eid = {L10},
        pages = {L10},
          doi = {10.1051/0004-6361/202449165},
archivePrefix = {arXiv},
       eprint = {2403.11923},
 primaryClass = {astro-ph.HE},
       adsurl = {https://ui.adsabs.harvard.edu/abs/2024A&A...684L..10M}
}

@article{Scargle13,
  author={Jeffrey D. Scargle and Jay P. Norris and Brad Jackson and James Chiang},
  title={Studies in Astronomical Time Series Analysis. VI. Bayesian Block Representations},
  journal={ApJ},
  volume={764},
  number={2},
  pages={167},
  url={http://stacks.iop.org/0004-637X/764/i=2/a=167},
  year={2013}
}

@ARTICLE{Stratta25,
       author = {{Stratta}, G. and {Nicuesa Guelbenzu}, A.~M. and {Klose}, S. and {Rossi}, A. and {Singh}, P. and {Palazzi}, E. and {Guidorzi}, C. and {Camisasca}, A. and {Bernuzzi}, S. and {Rau}, A. and {Bulla}, M. and {Ragosta}, F. and {Maiorano}, E. and {Paris}, D.},
        title = "{The Puzzling Long GRB 191019A: Evidence for Kilonova Light}",
      journal = {\apj},
         year = 2025,
        month = feb,
       volume = {979},
       number = {2},
          eid = {159},
        pages = {159},
          doi = {10.3847/1538-4357/ad9b7b},
archivePrefix = {arXiv},
       eprint = {2412.04059},
 primaryClass = {astro-ph.HE},
       adsurl = {https://ui.adsabs.harvard.edu/abs/2025ApJ...979..159S}
}

@ARTICLE{Svinkin16,
   author = {{Svinkin}, D.~S. and {Frederiks}, D.~D. and {Aptekar}, R.~L. and 
	{Golenetskii}, S.~V. and {Pal'shin}, V.~D. and {Oleynik}, P.~P. and 
	{Tsvetkova}, A.~E. and {Ulanov}, M.~V. and {Cline}, T.~L. and {Hurley}, K.},
    title = "{The Second Konus-Wind Catalog of Short Gamma-Ray Bursts}",
  journal = {ApJS},
archivePrefix = "arXiv",
   eprint = {1603.06832},
 primaryClass = "astro-ph.HE",
     year = 2016,
    month = may,
   volume = 224,
      eid = {10},
    pages = {10},
      doi = {10.3847/0067-0049/224/1/10},
   adsurl = {http://cdsads.u-strasbg.fr/abs/2016ApJS..224...10S}
}

@ARTICLE{Veres23,
       author = {{Veres}, P. and {Bhat}, P.~N. and {Burns}, E. and {Hamburg}, R. and {Fraija}, N. and {Kocevski}, D. and {Preece}, R. and {Poolakkil}, S. and {Christensen}, N. and {Bizouard}, M.~A. and {Dal Canton}, T. and {Bala}, S. and {Bissaldi}, E. and {Briggs}, M.~S. and {Cleveland}, W. and {Goldstein}, A. and {Hristov}, B.~A. and {Hui}, C.~M. and {Lesage}, S. and {Mailyan}, B. and {Roberts}, O.~J. and {Wilson-Hodge}, C.~A.},
        title = "{Extreme Variability in a Long-duration Gamma-Ray Burst Associated with a Kilonova}",
      journal = {\apjl},
         year = 2023,
        month = sep,
       volume = {954},
       number = {1},
          eid = {L5},
        pages = {L5},
          doi = {10.3847/2041-8213/ace82d},
archivePrefix = {arXiv},
       eprint = {2305.12262},
 primaryClass = {astro-ph.HE},
       adsurl = {https://ui.adsabs.harvard.edu/abs/2023ApJ...954L...5V}
}

@ARTICLE{Yang24,
       author = {{Yang}, Yu-Han and {Troja}, Eleonora and {O'Connor}, Brendan and {Fryer}, Chris L. and {Im}, Myungshin and {Durbak}, Joe and {Paek}, Gregory S.~H. and {Ricci}, Roberto and {Bom}, Cl{\'e}cio R. and {Gillanders}, James H. and {Castro-Tirado}, Alberto J. and {Peng}, Zong-Kai and {Dichiara}, Simone and {Ryan}, Geoffrey and {van Eerten}, Hendrik and {Dai}, Zi-Gao and {Chang}, Seo-Won and {Choi}, Hyeonho and {De}, Kishalay and {Hu}, Youdong and {Kilpatrick}, Charles D. and {Kutyrev}, Alexander and {Jeong}, Mankeun and {Lee}, Chung-Uk and {Makler}, Martin and {Navarete}, Felipe and {P{\'e}rez-Garc{\'\i}a}, Ignacio},
        title = "{A lanthanide-rich kilonova in the aftermath of a long gamma-ray burst}",
      journal = {\nat},
         year = 2024,
        month = feb,
       volume = {626},
       number = {8000},
        pages = {742-745},
          doi = {10.1038/s41586-023-06979-5},
archivePrefix = {arXiv},
       eprint = {2308.00638},
 primaryClass = {astro-ph.HE},
       adsurl = {https://ui.adsabs.harvard.edu/abs/2024Natur.626..742Y}
}

@ARTICLE{Yi25,
       author = {{Yi}, Shu-Xu and {Yorgancioglu}, Emre Seyit and {Xiong}, S. -L. and {Zhang}, S. -N.},
        title = "{Long pulse by short central engine: Prompt emission from expanding dissipation rings in the jet front of gamma-ray bursts}",
      journal = {Journal of High Energy Astrophysics},
         year = 2025,
        month = jul,
       volume = {47},
          eid = {100359},
        pages = {100359},
          doi = {10.1016/j.jheap.2025.100359},
       adsurl = {https://ui.adsabs.harvard.edu/abs/2025JHEAp..4700359Y}
}

@ARTICLE{Zhang06_nat,
       author = {{Zhang}, Bing},
        title = "{Astrophysics: A burst of new ideas}",
      journal = {\nat},
         year = 2006,
        month = dec,
       volume = {444},
       number = {7122},
        pages = {1010-1011},
          doi = {10.1038/4441010a},
archivePrefix = {arXiv},
       eprint = {astro-ph/0612614},
 primaryClass = {astro-ph},
       adsurl = {https://ui.adsabs.harvard.edu/abs/2006Natur.444.1010Z}
}

@ARTICLE{ICMART,
   author = {{Zhang}, B. and {Yan}, H.},
    title = "{The Internal-collision-induced Magnetic Reconnection and Turbulence (ICMART) Model of Gamma-ray Bursts}",
  journal = {ApJ},
archivePrefix = "arXiv",
   eprint = {1011.1197},
 primaryClass = "astro-ph.HE",
     year = 2011,
    month = jan,
   volume = 726,
      eid = {90},
    pages = {90},
      doi = {10.1088/0004-637X/726/2/90},
   adsurl = {http://adsabs.harvard.edu/abs/2011ApJ...726...90Z}
}

@ARTICLE{Zhong24,
       author = {{Zhong}, Shu-Qing and {Li}, Long and {Xiao}, Di and {Sun}, Hui and {Zhang}, Bin-Bin and {Dai}, Zi-Gao},
        title = "{The Very Early Soft X-Ray Plateau of GRB 230307A: Signature of an Evolving Radiative Efficiency in Magnetar Wind Dissipation?}",
      journal = {\apjl},
         year = 2024,
        month = mar,
       volume = {963},
       number = {1},
          eid = {L26},
        pages = {L26},
          doi = {10.3847/2041-8213/ad2852},
       adsurl = {https://ui.adsabs.harvard.edu/abs/2024ApJ...963L..26Z}
}

@ARTICLE{Dalessi25,
       author = {{Dalessi}, S. and {Veres}, P. and {Hui}, C.~M. and {Bala}, S. and {Lesage}, S. and {Briggs}, M.~S. and {Goldstein}, A. and {Burns}, E. and {Wilson-Hodge}, C.~A. and {Fletcher}, C. and {Roberts}, O.~J. and {Bhat}, P.~N. and {Bissaldi}, E. and {Cleveland}, W.~H. and {Giles}, M.~M. and {Godwin}, M. and {Hamburg}, R. and {Hristov}, B.~A. and {Kocevski}, D. and {Mailyan}, B. and {Malacaria}, C. and {Mukherjee}, O. and {Scotton}, L. and {von Kienlin}, A. and {Wood}, J.},
        title = "{Fermi-GBM Observations of GRB 230307A: An Exceptionally Bright Long-duration Gamma-ray Burst with an Associated Kilonova}",
      journal = {\apj},
         year = 2025,
        month = nov,
       volume = {994},
       number = {1},
          eid = {17},
        pages = {17},
          doi = {10.3847/1538-4357/ae0a1d},
archivePrefix = {arXiv},
       eprint = {2507.12637},
 primaryClass = {astro-ph.HE},
       adsurl = {https://ui.adsabs.harvard.edu/abs/2025ApJ...994...17D}
}

@ARTICLE{Lysenko25,
       author = {{Lysenko}, Alexandra L. and {Svinkin}, Dmitry S. and {Frederiks}, Dmitry D. and {Ridnaia}, Anna V. and {Tsvetkova}, Anastasia E. and {Ulanov}, Mikhail V.},
        title = "{The third Konus-Wind catalogue of short gamma-ray bursts}",
      journal = {\pasa},
         year = 2025,
        month = jun,
       volume = {42},
          eid = {e063},
        pages = {e063},
          doi = {10.1017/pasa.2025.10027},
archivePrefix = {arXiv},
       eprint = {2410.16896},
 primaryClass = {astro-ph.HE},
       adsurl = {https://ui.adsabs.harvard.edu/abs/2025PASA...42...63L}
}

@ARTICLE{Svinkin23_230304B,
       author = {{Kozyrev}, A.~S. and {Golovin}, D.~V. and {Litvak}, M.~L. and {Mitrofanov}, I.~G. and {Sanin}, A.~B. and {Mgns/Bepicolombo} and {Hend/Mars Odyssey Teams} and {Benkhoff}, J. and {Bepicolombo Team} and {Svinkin}, D. and {Fredericks}, D. and {Ridnaia}, A. and {Lysenko}, A. and {Cline}, T. and {Konus-Wind Team} and {Goldstein}, A. and {Briggs}, M.~S. and {Wilson-Hodge}, C. and {Burns}, E. and {Fermi Gbm Team} and {Bozzo}, E. and {Ferrigno}, C. and {INTEGRAL SPI-ACS Grb Team} and {Barthelmy}, S. and {Cummings}, J. and {Krimm}, H. and {Palmer}, D. and {Tohuvavohu}, A. and {Swift-Bat Team} and {Boynton}, W. and {Fellows}, C. and {Harshman}, K. and {Enos}, H. and {Starr}, R. and {Gardner}, A.~S. and {Grs-Odyssey Grb Team}},
        title = "{IPN triangulation of GRB 230304B}",
      journal = {GRB Coordinates Network},
         year = 2023,
        month = mar,
       volume = {33480},
        pages = {1},
       adsurl = {https://ui.adsabs.harvard.edu/abs/2023GCN.33480....1K}
}

\appendix
\section{Exponential fit of the four temporal properties}
\label{sec:exp_fit}
We used the following exponential models to describe the temporal behaviour of the PR, WTs, FWHMs, and peak energy.

\begin{align}
\mathrm{PR}(t) &= \mathrm{PR}_0\,\exp(-t/\tau_R), \label{eq:PR}\\
\mathrm{WT}(t) &= \frac{\tau_{\mathrm{WT}}}{N_0}\,\exp(t/\tau_{\mathrm{WT}}), \label{eq:WT}\\
 \mathrm{FWHM}(t) &= \mathrm{FWHM}_0\,\exp(t/\tau_{\mathrm{FWHM}}), \label{eq:FWHM}\\
E_p(t) &= E_{p,0}\,\exp\!\left(-t/\tau_{E_p}\right). \label{eq:Ep}
\end{align}
$PR_0$, $\rm FWHM_0$, and $E_{p,0}$ are the initial values at $t=0$, and $\tau_R$, $\tau_{\rm WT}$, $\tau_{\mathrm{FWHM}}$, $\tau_{\rm E_p}$ the characteristic evolution timescales. $N_0$ is the number of elementary bunches of energy in the toy model defined in \citetalias{Maccary26} used to model the WTs evolution.

These relations were modelled using the D'Agostini likelihood, in a similar way as for the modelling of the $E_{\rm p,i}-E_{\rm iso}$ previously described, and the dispersion of these relations are described by $\sigma_R$, $\sigma_{\rm WT}$, $\sigma_{\rm FWHM}$, and $\sigma_{\rm E_p}$, respectively. The evolution of these four properties are displayed in Fig.~\ref{fig:evolution_three} and Fig.~\ref{fig:evolution_six_others}.
\begin{figure*}[h!]
\centering
\includegraphics[width=\linewidth]{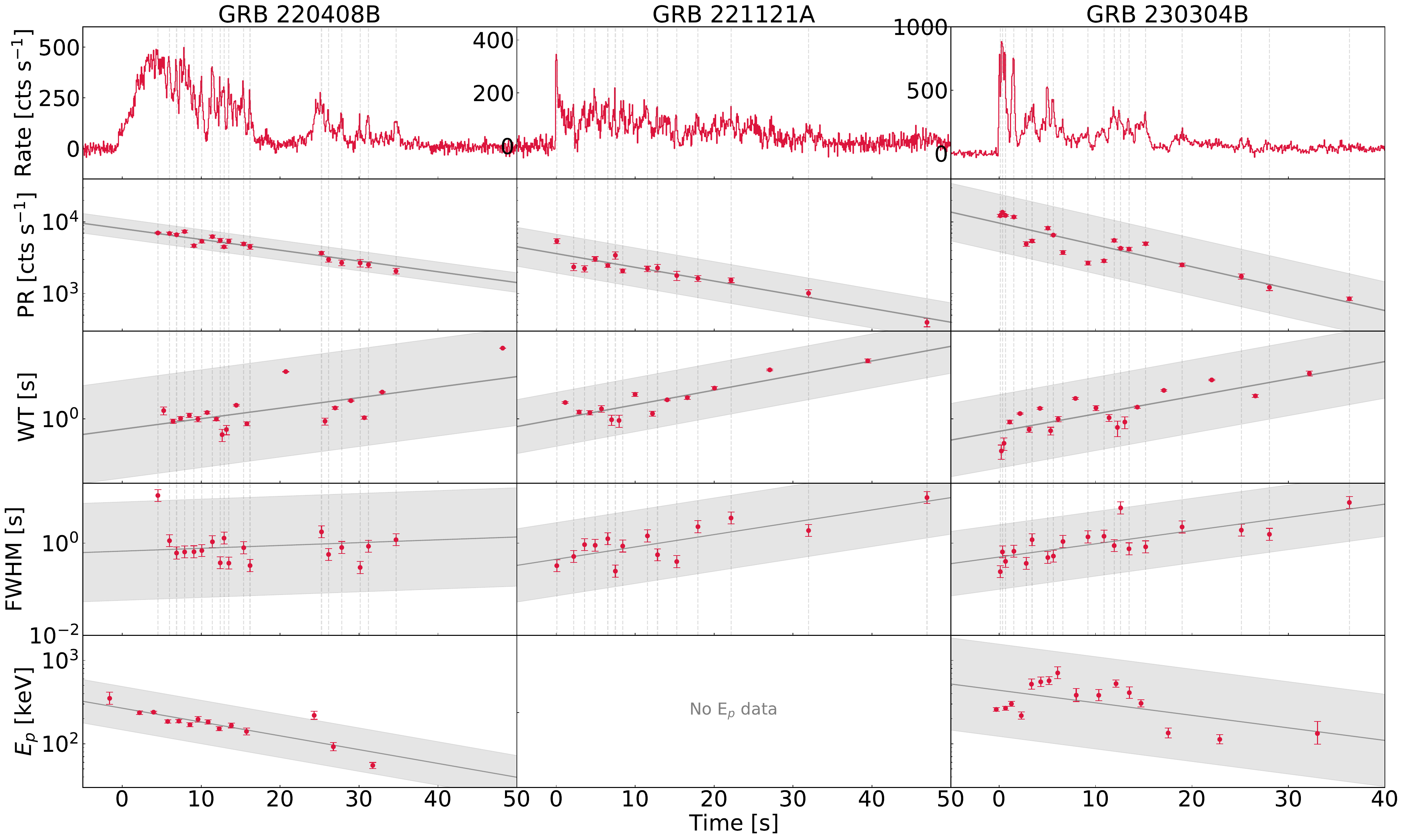}

\includegraphics[width=\linewidth]{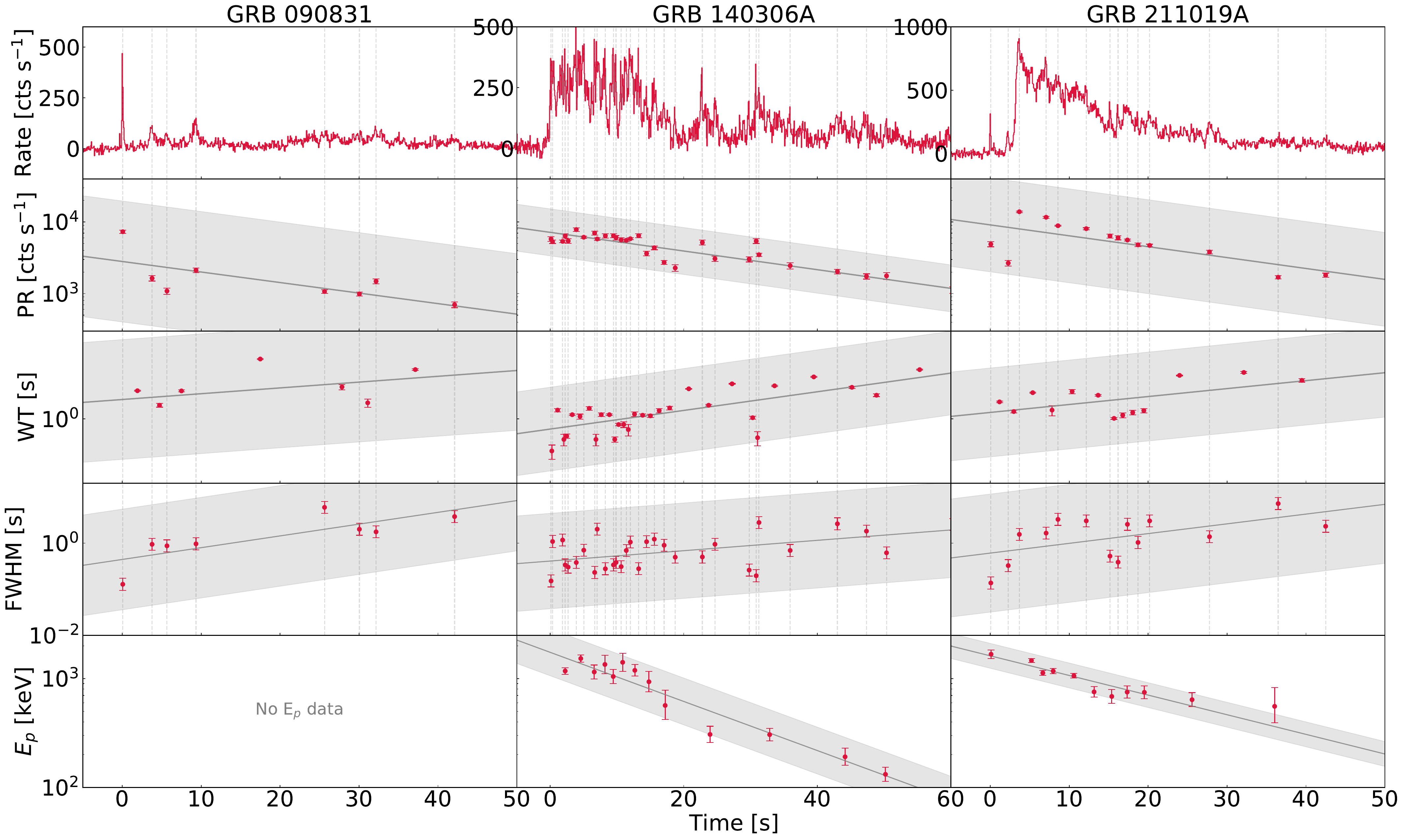}
\caption{Temporal evolution of the PR, WTs, FWHM, and $E_{\mathrm{p}}$ for the six remaining GRBs of the sample $S_0$. The grey solid lines represent the best fits of the linear relation and the shaded grey regions represent the 3$\sigma$ confidence intervals of the fits.}
\label{fig:evolution_six_others}
\end{figure*}
The parameters of the fit are given in Table~\ref{tab:fit_results_3ev}, and Table~\ref{tab:Ep_fit_results} for the fit of the peak energy.

\begin{table*}[ht]
\centering
\caption{Best-fit parameters for each GRB of the $S_0$ sample for the three different fits (WTs, FWHM, and PR).}
\begin{tabular}{lcccccccccc}
\hline
GRB & $\tau_{\mathrm{WT}}$ (s) & $N_0$ & $\sigma_{\mathrm{WT}}$ &
$\tau_{\mathrm{FWHM}}$ (s) & $\mathrm{FWHM}_0$ (s) & $\sigma_{\mathrm{FWHM}}$ &
$\tau_{R}$ (s) & $\mathrm{PR}_0$ ($\times10^3$ cts/s) & $\sigma_{R}$ \\
\hline
\\
GRB\,090831 & $37^{+270}_{-24}$ & $15^{+64}_{-8}$ & $0.93^{+0.74}_{-0.36}$ & $17^{+28}_{-7}$ & $0.44^{+0.67}_{-0.27}$ & $0.84^{+0.73}_{-0.36}$ & $30^{+75}_{-17}$ & $2.8^{+2.7}_{-1.4}$ & $0.65^{+0.59}_{-0.25}$ \\
GRB\,140306A & $23.0^{+7.0}_{-4.0}$ & $37^{+9}_{-7}$ & $0.65^{+0.19}_{-0.13}$ & $39^{+30}_{-12}$ & $0.41^{+0.17}_{-0.12}$ & $0.79^{+0.24}_{-0.17}$ & $33^{+5}_{-4}$ & $7.2^{+0.9}_{-0.8}$ & $0.25^{+0.07}_{-0.05}$ \\
GRB\,150510A & $14.7^{+5.3}_{-3.0}$ & $25^{+7}_{-5}$ & $0.59^{+0.23}_{-0.14}$ & $21^{+8}_{-5}$ & $0.24^{+0.08}_{-0.06}$ & $0.73^{+0.21}_{-0.15}$ & $20^{+2}_{-2}$ & $16^{+2}_{-2}$ & $0.26^{+0.07}_{-0.05}$ \\
GRB\,170527A & $16.4^{+3.9}_{-2.6}$ & $31^{+5}_{-4}$ & $0.42^{+0.13}_{-0.09}$ & $21^{+12}_{-6}$ & $0.43^{+0.20}_{-0.14}$ & $0.68^{+0.24}_{-0.17}$ & $22^{+5}_{-3}$ & $17.2^{+3.0}_{-2.6}$ & $0.31^{+0.09}_{-0.07}$ \\
GRB\,190415A & $18.3^{+6.3}_{-3.7}$ & $30^{+8}_{-6}$ & $0.62^{+0.21}_{-0.14}$ & $23^{+14}_{-6}$ & $0.32^{+0.16}_{-0.10}$ & $0.79^{+0.27}_{-0.19}$ & $22^{+3}_{-2}$ & $7.9^{+1.1}_{-1.1}$ & $0.29^{+0.10}_{-0.07}$ \\
GRB\,211019A & $27^{+68}_{-12}$ & $20^{+28}_{-7}$ & $0.69^{+0.36}_{-0.20}$ & $21^{+54}_{-10}$ & $0.61^{+0.73}_{-0.33}$ & $0.98^{+0.52}_{-0.30}$ & $29^{+36}_{-11}$ & $9^{+4}_{-3}$ & $0.50^{+0.25}_{-0.14}$ \\
GRB\, 220408B & $20.4^{+14.6}_{-5.7}$ & $33^{+13}_{-9}$ & $0.76^{+0.30}_{-0.18}$ & $71^{+391}_{-40}$ & $0.67^{+0.35}_{-0.27}$ & $0.82^{+0.33}_{-0.22}$ & $29^{+3}_{-2}$ & $8.1^{+0.6}_{-0.6}$ & $0.11^{+0.04}_{-0.03}$ \\
GRB\,221121A& $14.7^{+6.0}_{-3.4}$ & $15^{+4}_{-3}$ & $0.42^{+0.24}_{-0.13}$ & $16^{+12}_{-5}$ & $0.45^{+0.28}_{-0.17}$ & $0.61^{+0.36}_{-0.23}$ & $23^{+6}_{-4}$ & $3.6^{+0.6}_{-0.5}$ & $0.21^{+0.12}_{-0.07}$ \\
GRB\,230304B & $12.3^{+5.9}_{-3.0}$ & $22^{+7}_{-5}$ & $0.56^{+0.26}_{-0.15}$ & $15^{+9}_{-4}$ & $0.50^{+0.22}_{-0.15}$ & $0.54^{+0.26}_{-0.18}$ & $14^{+3}_{-2}$ & $9.7^{+2.1}_{-1.8}$ & $0.34^{+0.14}_{-0.08}$ \\
\\
\hline
\end{tabular}
\label{tab:fit_results_3ev}
\end{table*}

\begin{table*}[ht]
\centering
\caption{Best-fit results of the exponential evolution of $E_{\mathrm{p}}$ with time for each GRB of the $S_0$ sample.}
\begin{tabular}{lcccc}
\hline
GRB & $\tau_{E_{\mathrm{p}}}$ (s) & $E_{\mathrm{p,0}}$ (keV) & $\sigma_{E_{\mathrm{p}}}$ & $\chi^2$/d.o.f. \\
\hline
\\
GRB\,140306A & $19.2^{+3.5}_{-2.7}$ & $1749^{+393}_{-299}$ & $0.21^{+0.15}_{-0.09}$ & $9.2/10$ \\
GRB\,150510A & $16.3^{+2.9}_{-2.0}$ & $1759^{+419}_{-329}$ & $0.18^{+0.17}_{-0.09}$ & $6.9/7$ \\
GRB\,170527A & $19.2^{+9.4}_{-4.7}$ & $1150^{+442}_{-313}$ & $0.39^{+0.18}_{-0.11}$ & $17/18$ \\
GRB\,190415A & $15.2^{+1.7}_{-1.4}$ & $1526^{+254}_{-198}$ & $0.15^{+0.13}_{-0.06}$ & $4.6/9$ \\
GRB\,211019A & $24.1^{+9.3}_{-4.5}$ & $1619^{+202}_{-232}$ & $0.09^{+0.11}_{-0.06}$ & $8.2/8$ \\
GRB\,220408B & $26.2^{+13.2}_{-6.7}$ & $268^{+59}_{-48}$ & $0.24^{+0.14}_{-0.08}$ & $11.4/11$ \\
GRB\,230304B & $29^{+44}_{-12}$ & $436^{+162}_{-112}$ & $0.48^{+0.22}_{-0.13}$ & $12.9/14$ \\
\\
\hline
\end{tabular}
\label{tab:Ep_fit_results}
\end{table*}


\section{Statistical tests}
\label{sec:stat_test}
Figure~\ref{fig:likelihood_S0_GRBs} represents the likelihood values of the $S_0$ sample testing the compatibility of these GRBs to the $E_{\rm p,i}-E_{\rm iso}$ relation for type-II GRBs. \begin{figure*}[h!]
    \centering
\includegraphics[width=0.49\textwidth]{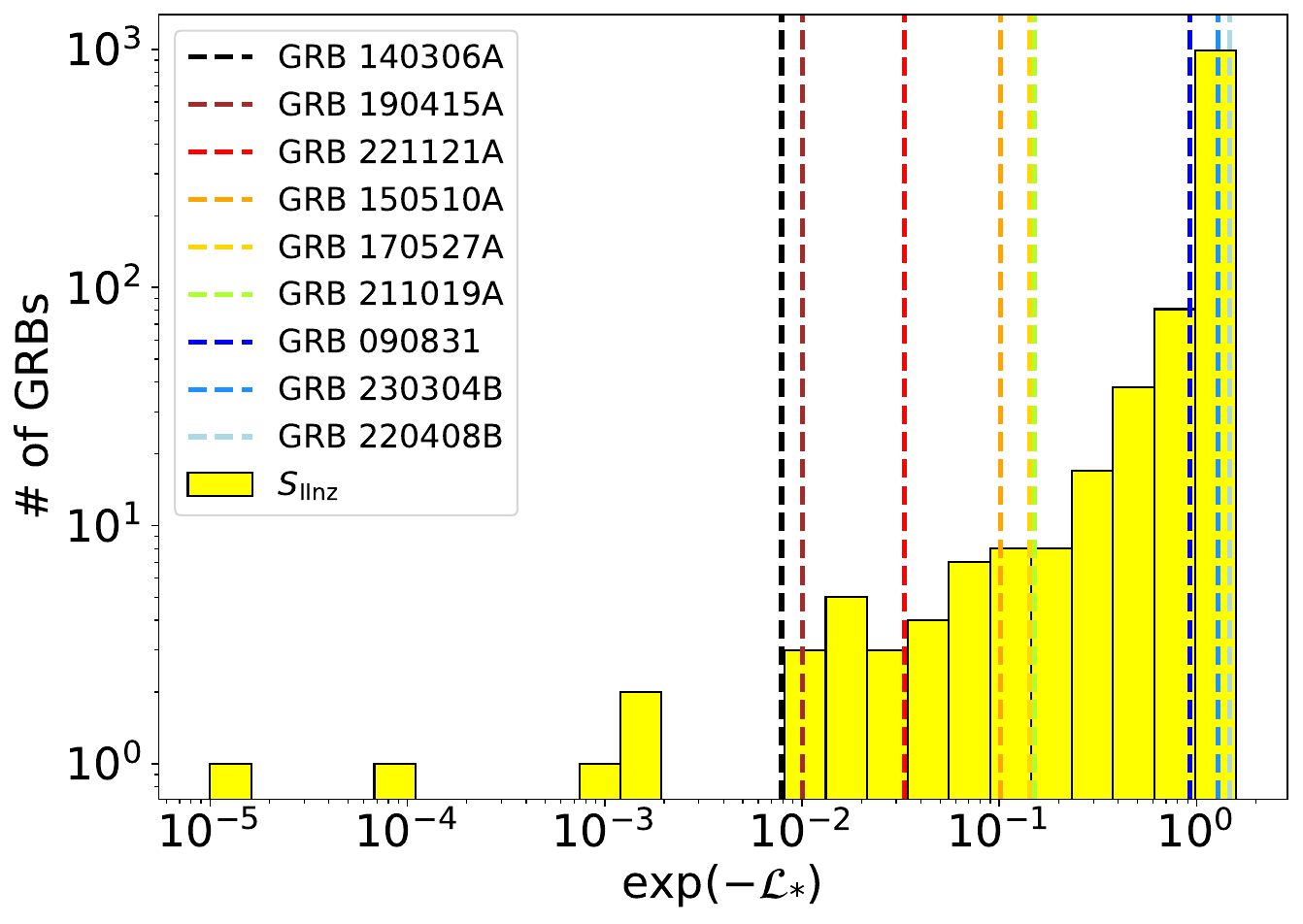}
    \includegraphics[width=0.49\textwidth]{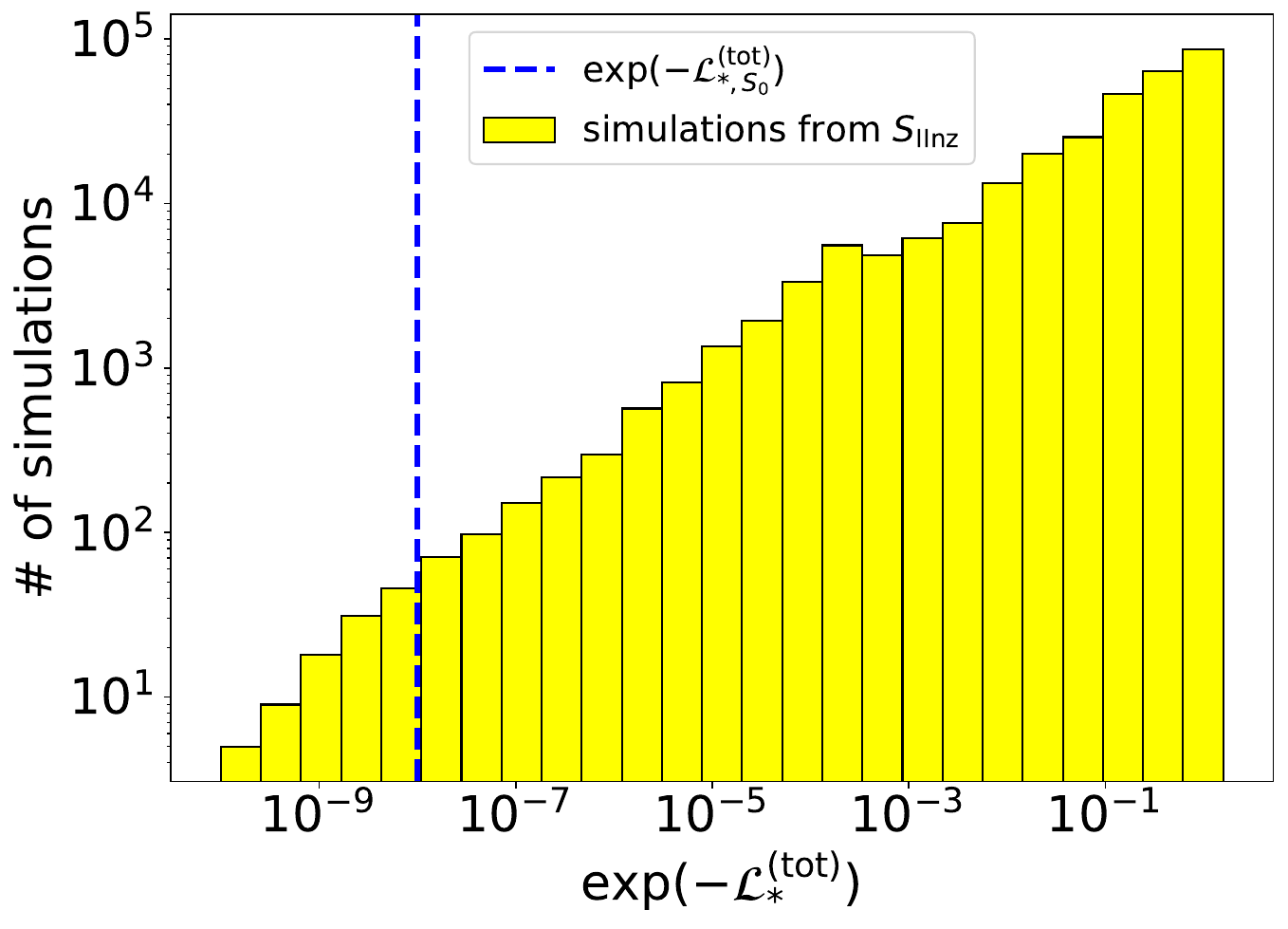}
    \caption{\textit{Left}: distribution of likelihoods (given by $\exp(-\mathcal{L_*})$, $\mathcal{L_*}$ being the negative log-likelihood) for GRBs from sample $S_{\rm IInz}$. Vertical lines show the values of $S_0$ candidates. \textit{Right}: histogram of $\exp{(-\mathcal{L_*}^{\rm (tot)})}$, where $\mathcal{L_*}^{\rm (tot)}=\sum_{i=1}^{9} \mathcal{L_*}^{(i)}$ obtained from $10^{6}$ realisations of random selection of 9 GRBs from $S_{ \rm IInz}$ (Sec.~\ref{sec:EpiEiso}). The vertical blue line shows the value obtained from the sample $S_0$, $\exp{(-{\cal L}_{*,S_0}^{\rm (tot)})}$. The fraction of simulations for which is it ${\cal L}_{*}^{\rm (tot)} \ge {\cal L}_{*,S_0}^{\rm (tot)}$ is the joint probability that all 9 GRBs from $S_0$ are consistent with the $E_{\rm p,i}-E_{\rm iso}$ for type-II GRBs. This fraction turns out to be $1.2 \times 10^{-4}$, equivalent to $3.7 \sigma$ (Gaussian).
    }
    \label{fig:likelihood_S0_GRBs}
\end{figure*}
Table~\ref{tab:stats_tests.} reports the results of the statistical tests.

\begin{table}[h!]
\centering
\caption{Probabilities for a set of $H_0$ hypotheses, each of which assumes a maximum value of the number of outliers of the $E_{\rm p,i}-E_{\rm iso}$ relation in the $S_0$ sample (Sect.~\ref{sec:EpiEiso}).}
\label{tab:h0_tests}
\begin{tabular}{ccc}
\hline
$N_{\rm out}$  & $p$-value & $\sigma$ Gaussian\\
\hline
0 & $1.24 \times 10^{-4}$ & $3.7$ \\
1 & $8.21 \times 10^{-4}$ & $3.1$ \\
2 & $8.46 \times 10^{-3}$ & $2.4$ \\
3 & $2.10 \times 10^{-2}$ & $2.0$ \\
4 & $4.40 \times 10^{-2}$ & $1.7$ \\
\hline
\end{tabular}
\label{tab:stats_tests.}
\end{table}

\section{Correlation tests}
\label{sec:corr_tests}

To further assess whether the temporal and spectral properties of the prompt emission evolve during the burst, we tested for correlations between time, described by the peak time of the pulses detected by {\sc mepsa}, and the four properties studied in \citetalias{Maccary26}, i.e. PR, WTs, pulse FWHM, and, when available, spectral peak energy $E_p$.
For each observable, we evaluated the presence of a linear trend with time using a standard correlation test. Since all four properties show an exponential evolution, the analysis was performed in logarithmic space for the dependent variables. A Pearson test, testing for linear correlation, was performed. We computed the corresponding correlation coefficient and associated $p$-value, which quantifies the probability of obtaining the observed correlation by chance under the null hypothesis of no temporal trend. The result of this analysis is shown in Figure~\ref{fig:corr_test}.
\begin{figure}[htpb]
    \centering
    \includegraphics[width=\linewidth]{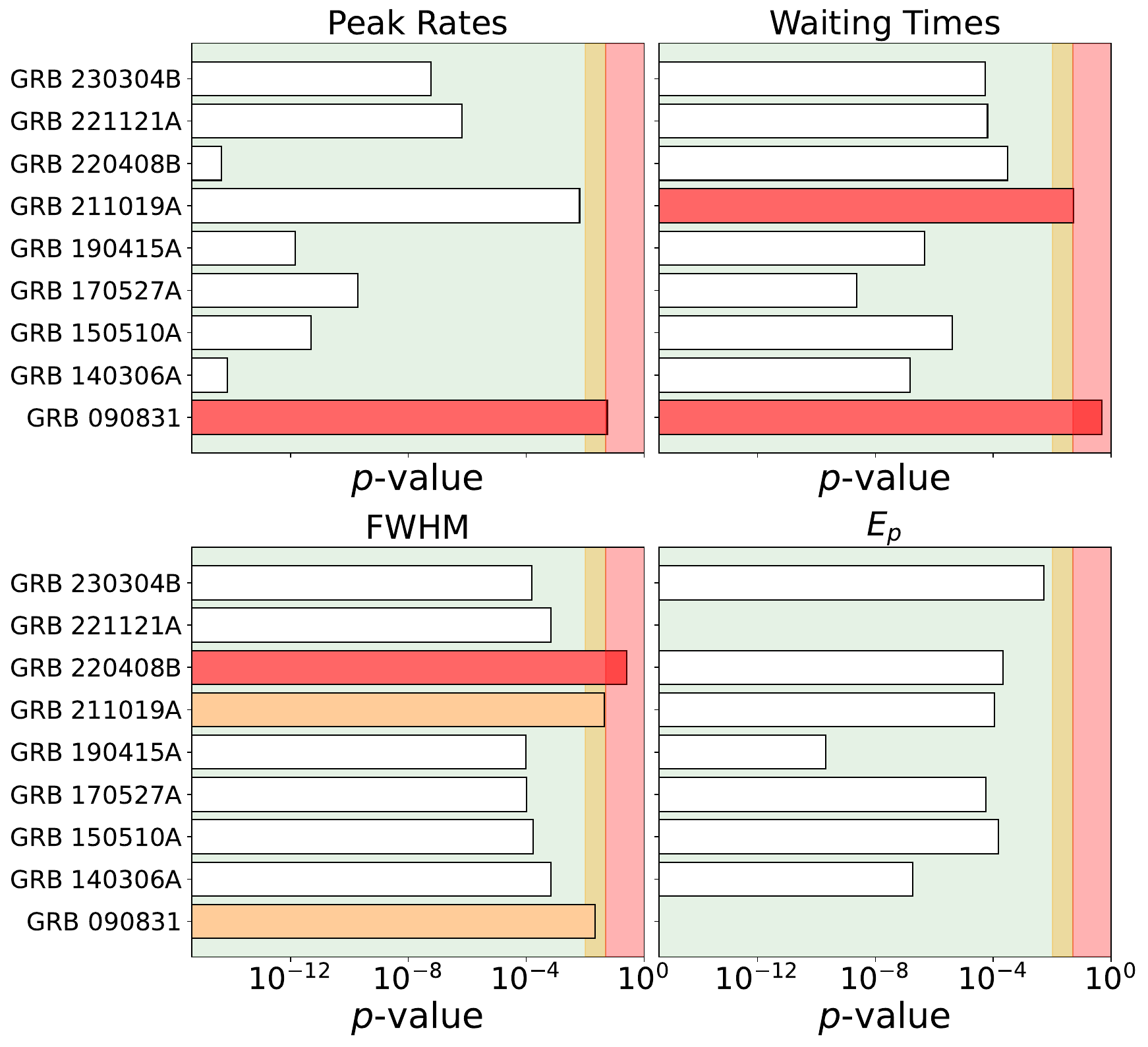}
    \caption{\textit{From top to bottom panels, left to right}: $p$-values from Pearson correlation tests for the PR, WTs, pulse FWHM, and, when available, the spectral peak energy $E_p$. Horizontal bars indicate the $p$-values associated with each correlation test. Green-shaded regions (corresponding to white bars) denote $p$-values below $0.01$, for which the null hypothesis of no temporal trend is rejected. Orange bars correspond to $p$-values between $0.01$ and $0.05$, indicating marginal detections of temporal trends. Red bars indicate $p$-values greater than $0.05$, for which the null hypothesis of no temporal evolution cannot be rejected.}
    \label{fig:corr_test}
\end{figure} Several GRBs (GRB\,140306A, GRB\,150510A, GRB\,170527A, GRB\,190415A, GRB\,230304B) have all four properties correlated with time (or at least three when the time-integrated spectrum is not available, as for GRB\,221121A). In some cases (GRB 211019A, GRB\,220408B), for one or two properties the correlation cannot be established. Only for GRB\, 090831, the faintest GRB considered in sample $S_0$, no correlation can be firmly established. 

\section{Redshift limiting values from the violation of the Amati relation of type-II GRBs}
\label{sec:ul_redshfit}
The $E_{\rm p,i}-E_{\rm iso}$ plane can be used to estimate the redshift below which a given GRB in the sample $S_0$ becomes incompatible (at $3$ or $2\sigma$ confidence level) with the $E_{\rm p,i}-E_{\rm iso}$ followed by the bulk of collapsar GRBs. This value is obtained by computing the redshift at which the track formed by a given GRB intersects the $3\sigma_{\rm int}$ line (resp. $2\sigma_{\rm int}$ line) of the $E_{\rm p,i}-E_{\rm iso}$ relation. The results are reported in Table~\ref{tab:grb_redshift}. Although this quantity does not directly provide an upper limit on the GRB redshift (as several type-I GRBs are very close to the Amati relation  of type-II GRBs) GRBs, this information could be used to ease the research of an associated host galaxy. In particular, for GRB\,090831, GRB\,220408B, GRB\,230304B, the obtained limits would place these GRBs at very close distances. However, we emphasise that these values should not be taken as direct estimates of the burst redshift. Indeed, as for instance is the case for GRB\,230307A, a type-I GRB can perfectly lie within the 3 $\sigma$ relation of type-II GRBs. 
\begin{table}[ht]
\centering
\caption{Redshift values below which each GRB becomes incompatible with the Amati relation for type-II GRBs at $2$ and $3\sigma$ confidence levels.}
\label{tab:grb_redshift}
\begin{tabular}{lcc}
\hline
GRB & $z_{3\sigma}$ & $z_{2\sigma}$ \\
\hline
 GRB\,090831 & $0.050$ & $0.14$  \\
 GRB\,140306A & $0.88$ & $-$  \\
 GRB\,150510A & $0.19$  & $0.97$  \\
 GRB\,170527A & $0.16$ & $0.65$  \\
 GRB\,190415A & $0.67$ & $-$ \\ 
 GRB\,211019A & $0.16$ & $0.62$ \\ 
 GRB\,220408B & $0.021$ & $0.054$ \\
 GRB\,221121A & $0.34$ & $-$ \\
 GRB\,230304B & $0.031$ & $0.083$ \\
\hline
\end{tabular}

\end{table}

\section{The impact of using different MVT metrics}
\label{sec:impact_MVT_metrics}
Various studies have focused on computing the MVT. Notably, \citet{MacLachlan12,MacLachlan13}, and \citet{Golkhou14,Golkhou15} developed non-parametric, wavelet-based measures of intrinsic variability. These methods are designed to study the temporal power as a function of timescale in a robust and unbiased manner, without relying on pulse fitting procedures or specific assumptions regarding pulse shape. The resulting MVT values are found to be closely related to the shortest rise times in GRB light curves. In contrast, the MVT defined in \citet{Camisasca23} and in \citet{Maccary25}, computed as the FWHM of the shortest pulse in the light curve, is found to be about $5-6$ times longer than the MVT derived from the aforementioned studies. However, this discrepancy does not impact the conclusions drawn from Figure~\ref{fig:mvt_t90}. In fact, adopting alternative MVT definitions would essentially result into a leftward, roughly uniform, shift of all data points. Consequently, the relative positioning of the $9$ candidates against the bulk population remains fundamentally unchanged. This consistency is illustrated in Figure~\ref{fig:M13_vs_FWHM_min}, which compares the MVT-$T_{90}$ plane using the GRB sample from \citet{MacLachlan13} (left panel) with values sourced from both that catalogue and \citet{Maccary25} (right panel).

\begin{figure}[h!]
    \centering
    \includegraphics[width=\linewidth]{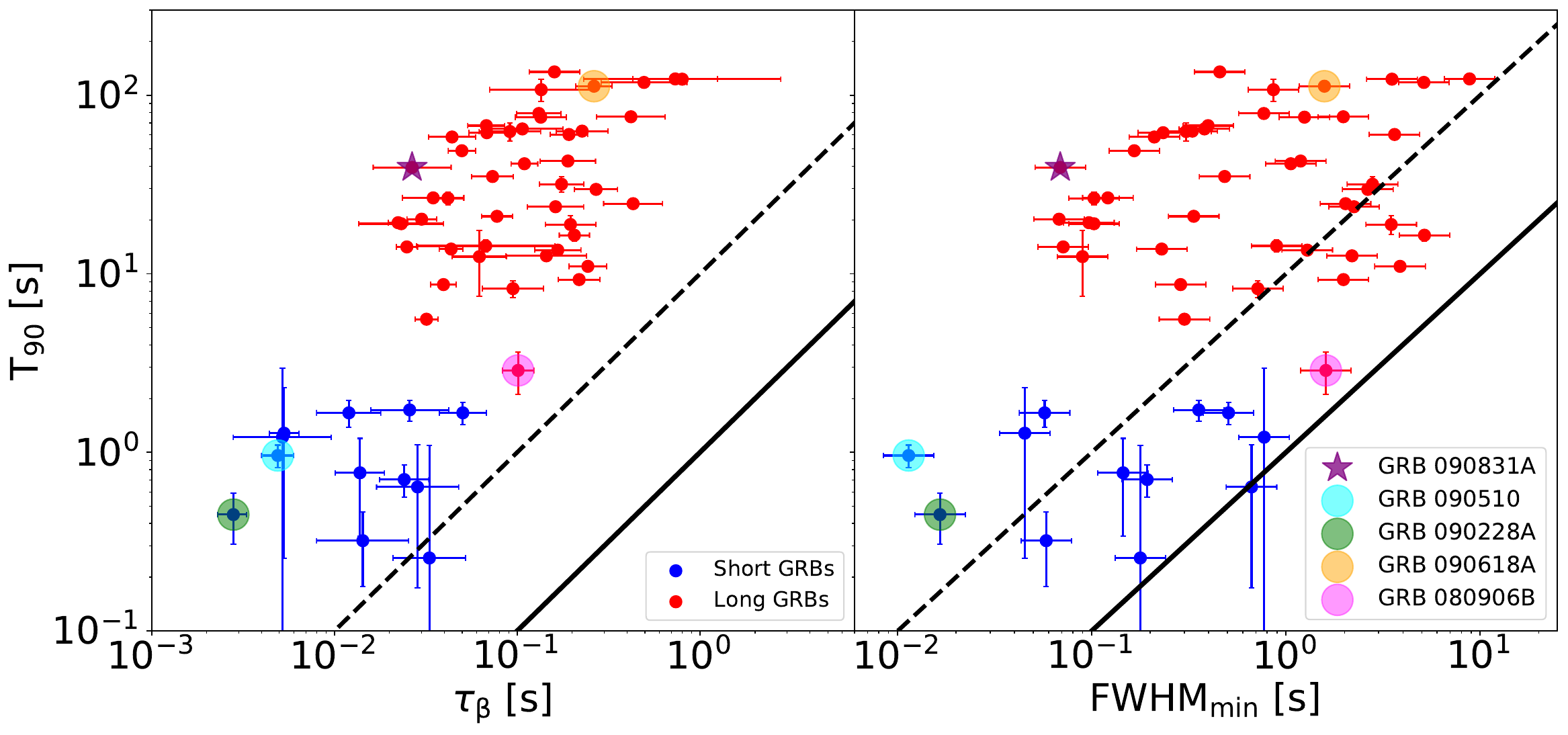}
    \caption{\textit{Left:} MVT-$T_{90}$ plane, with the MVT (denoted as $\tau_{\beta}$) values taken from \citet{MacLachlan13}, using their GRB sample. \textit{Right:} MVT-$T_{90}$ plane for the same bursts using the MVT values (denoted as ${\rm FWHM_{min}}$) from \citet{Maccary25}. Solid and dashed lines show equality and $T_{90} = 10 \times$ MVT. Some bursts are shown with coloured filled circles with the only purpose of highlighting the corresponding location in each plot. GRB\,090831A, which belongs to the $S_0$ sample, is identified with a purple star. Evidently, the relative positioning of these bursts remains essentially invariant, regardless of the MVT metric employed.}
    \label{fig:M13_vs_FWHM_min}
\end{figure}

In spite of subtle differences between the two sets and a significant overlap between the two populations of long and short GRBs, the latter consistently exhibit lower MVT values than the former. Notably, the MVT values of the two populations from \citet{MacLachlan13} (left panel) appear to be slightly better separated than the corresponding $\rm FWHM_{\rm min}$ values (right panel). Ultimately, for the scope of this study, both MVT definitions lend support to the same conclusions.

\end{document}